\documentclass[aps,prd,showpacs,nobibnotes,twocolumn,preprintnumbers,
nobalancelastpage,superscriptaddress]{revtex4}
\usepackage{latexsym}
\usepackage{dcolumn}
\usepackage{bm}
\usepackage{natbib}
\input{epsf}
\newcommand{\be}{\begin{equation}}
\newcommand{\ee}{\end{equation}}
\newcommand{\ba}{\begin{eqnarray}}
\newcommand{\ea}{\end{eqnarray}}
\newcommand{\bi}{\begin{itemize}}
\newcommand{\ei}{\end{itemize}}
\newcommand{\bfi}{\begin{figure}[!t]
\epsfxsize=9cm
\epsffile}
\newcommand{\bfig}{\begin{figure*}[t]
\epsfxsize=15cm
\epsffile}
\newcommand{\efi}{\end{figure}}
\newcommand{\efig}{\end{figure*}}
\newcommand{\no}{\nonumber}
\newcommand{\la}{\lesssim}
\newcommand{\ga}{\gtrsim}

\begin{document}
\title{Peculiar velocity decomposition, redshift space distortion and velocity
  reconstruction in redshift surveys. II. Dark matter velocity statistics}
\author{Yi Zheng}
\affiliation{Key Laboratory for Research in Galaxies and Cosmology, Shanghai
  Astronomical Observatory, 80 Nandan Road, Shanghai, 200030,
  China}
\author{Pengjie Zhang}
\email[Email me at: ]{zhangpj@sjtu.edu.cn}
\affiliation{Center for Astronomy and Astrophysics, Department of
  Physics and Astronomy, Shanghai Jiao Tong
  University, 955 Jianchuan road, Shanghai, 200240}
\affiliation{Key Laboratory for Research in Galaxies and Cosmology, Shanghai
  Astronomical Observatory, 80 Nandan Road, Shanghai, 200030,
  China}
\author{Yipeng Jing}
\affiliation{Center for Astronomy and Astrophysics, Department of
  Physics and Astronomy, Shanghai Jiao Tong
  University, 955 Jianchuan road, Shanghai, 200240}
\author{Weipeng Lin}
\affiliation{Key Laboratory for Research in Galaxies and Cosmology, Shanghai
  Astronomical Observatory, 80 Nandan Road, Shanghai, 200030,
  China}
\author{Jun Pan}
\affiliation{National Astronomical Observatories, Chinese Academy of Sciences,
20A Datun Road, Chaoyang District, Beijing 100012, P. R. China}


\begin{abstract}
Massive spectroscopic redshift surveys open a promising window to accurately
measure peculiar velocity at cosmological distances through redshift space distortion (RSD).
In Paper I \cite{paperI}  of this series of work, we proposed decomposing
peculiar velocity into three eigenmodes (${\bf  v}_\delta$, ${\bf
  v}_S$ and ${\bf  v}_B$) in order to facilitate the RSD modeling and
peculiar velocity reconstruction.  In the current paper we measure the
dark matter RSD-related statistics of the velocity eigenmodes through a
set of N-body simulations. These statistics  include the velocity
power spectra, correlation functions, one-point probability
distribution functions, cumulants, and the
damping functions describing the Finger of God effect.  We have
carried out a number of tests to quantify possible numerical artifacts
in these measurements and have confirmed  that these numerical
artifacts are under control. Our major findings are as follows: (1) The
power spectrum measurement shows that these velocity components
have distinctly different spatial distribution and redshift evolution,
consistent with predictions in Paper I. In particular, we measure the
window function $\tilde{W}(k,z)$. $\tilde{W}$ describes the impact of
nonlinear evolution on the ${\bf
  v}_\delta$-density relation. We confirm that the approximation
$\tilde{W}=1$ can induce a significant systematic 
error of $O(10\%)$ in RSD cosmology. We demonstrate that $\tilde{W}$ can be accurately
described by a simple fitting formula with one or two free parameters.
(2) The correlation function
measurement shows that the correlation length is $O(100)$, $O(10)$, and
$O(1)$ Mpc for  ${\bf v}_\delta$,
${\bf v}_S$, and ${\bf v}_B$ respectively.  These correlation lengths
determine where we can treat the velocity fields as spatially
uncorrelated. Hence, they are important properties in RSD modeling.  (3)
The velocity probability distribution functions and cumulants quantify
non-Gaussianities of the velocity fields. We confirm speculation in
Paper I that ${\bf v}_\delta$ is largely Gaussian, but with
non-negligible non-Gaussianity.  We confirm that  ${\bf v}_B$ is
significantly non-Gaussian. We also measure the damping
functions. Despite the observed non-Gaussianities, the damping
functions and hence the Finger of God effect are all well approximated as Gaussian ones at scales of
interest.
\end{abstract}
\maketitle

\section{Introduction}
Redshift space distortion (RSD) is emerging as a powerful probe of
dark energy and gravity at cosmological scales
\cite{Jackson72,Sargent77,Peebles80,Kaiser87,Peacock94,Ballinger96,Peacock01,Tegmark02,
Tegmark04,Amendola05, Linder05,Yamamoto05,
Zhang07c,Guzzo08,Wang08,Percival09,Song09,White09,Song10,Wang10,Blake11b,Samushia12,Blake12,Reid12,Tojeiro12,Jain08,Linder08,Reyes10,Cai12,Gaztanaga12,Jennings12,Li12,BigBOSS,Euclid}.
 However, RSD modeling is highly complicated (e.g., \cite{Scoccimarro04,Seljak11,
  Okumura12,Okumura12b,Bonoli09,Desjacques10} and references therein),
due to entangled nonlinearities in the velocity and density fields and
in the real space-redshift space mapping. 

RSD is induced by peculiar motion. Naturally the
first step to understand RSD is to understand the peculiar velocity
${\bf v}$. It turns out that different components of ${\bf v}$ affect RSD
in different ways. Motivated by it, we proposed a unique
decomposition of ${\bf v}$ into three eigenmodes (Paper I, \cite{paperI}). The three
components are not only uniquely defined mathematically, but also have
unique physical meanings and have different impacts on RSD. We find
that this decomposition indeed facilitates the RSD modeling. It helps
us in arriving at a new RSD formula. It also helps us in proposing new
approaches to reconstruct three-dimensional peculiar velocity at cosmological
distances through spectroscopic redshift surveys.

Paper I outlined the basic methodology, but left most quantitative
analysis for future studies. The present paper (Paper II) is the second paper of
the series, focusing on the RSD-related dark matter velocity statistics. Future
works will extend to the halo velocity statistics and eventually to the
galaxy velocity statistics. We will evaluate the accuracy of the proposed RSD
formula for dark matter and galaxies. We will also  investigate the proposed
velocity reconstruction techniques.

This paper is organized as follows. \S \ref{sec:theory} briefly describes the
three velocity components. \S \ref{sec:simulation} describes the
simulations and the velocity assignment method used for the
analysis. Numerical artifacts are quantified through several tests detailed
in the Appendix and are proven to be under control. \S \ref{sec:results} shows
the results of relevant velocity statistics. We summarize and discuss
in \S \ref{sec:summary}.


\section{Theory basics}
\label{sec:theory}
Paper I proposed decomposing the velocity ${\bf v}$ into three
eigenmodes (${\bf v}_\delta$, ${\bf v}_S$, and ${\bf v}_B$). This
decomposition allows us to conveniently derive the leading-order RSD corrections in
the redshift space power spectrum. For the dark matter redshift space power spectrum, 
\ba
\label{eqn:RD4}
P_{\delta\delta}^s(k,u)&\simeq&
  \left(P_{\delta\delta}(k)\left(1+f\tilde{W}(k)u^2\right)^2 \right.\\
&&\left. +u^4P_{\theta_S\theta_S}(k)+C_G(k,u)+C_{NG,3}(k,u)\right)\no \\
&\times&
   D^{\rm FOG}(ku)\ . \no
\ea
Here, $u\equiv k_z/k$. The FOG effect is described by the damping function
\be
D^{\rm FOG}(ku)=D_{\delta}^{\rm FOG}(ku) D_S^{\rm FOG}(ku) D_B^{\rm
  FOG}(ku)\ .
\ee
We predict in paper I that ${\bf v}_B$ only contributes to the Finger of God (FOG)
effect through $D^{\rm FOG}_B$. ${\bf v}_S$ causes both an enhancement
($P_{\theta_S\theta_S}$) and a damping ($D^{\rm FOG}_S$).  All other
terms in Eq. (\ref{eqn:RD4}) are contributed by ${\bf
  v}_\delta$. For more
details, refer to Paper I. Here we just summarize the basic results of the
three velocity components, with terms in Eq. (\ref{eqn:RD4}) explained. 

 ${\bf v}_\delta$ is irrotational ($\nabla \times {\bf
    v}_\delta=0$) and is completely correlated with the underlying
  density field $\delta$. The relation in Fourier space is
\ba
\label{eqn:thetam}
\theta_\delta({\bf k})&=&\delta({\bf k})W({\bf k})\ , \\
\ W({\bf k})&=&W(k)=\frac{P_{\delta\theta}(k)}{P_{\delta\delta}(k)}\
. \no
\ea
Here,  $\theta_\delta\equiv -\nabla\cdot {\bf v}_\delta/H$ and  $\theta\equiv -\nabla \cdot {\bf
  v}/H$. $P_{AB}$ is the power spectrum between the field $A$ and the
field $B$, defined as $\left\langle A({\bf k}) B({\bf k}^\prime)\right\rangle \equiv
(2\pi)^3\delta_{3D}({\bf k}+{\bf k}^\prime)P_{AB}({\bf k})$. We often use the notation $\Delta^2_{AB}\equiv k^3P_{AB}/(2\pi^2)$, which enters into the ensemble average $\langle A({\bf
  x})B({\bf x})\rangle=\int \Delta^2_{AB}(k)dk/k$. If $A$ and $B$ are
vector fields, $\langle AB\rangle\rightarrow \langle {\bf A}\cdot{\bf
  B}\rangle$.

In the limit $k\rightarrow 0$, $W\rightarrow f\equiv d\ln
D/d\ln a$ for the matter field \footnote{For the galaxy field,
  $W_g\rightarrow\beta\equiv f/b_g$ in the limit $k\rightarrow
  0$. Refer to paper I for detailed discussions. }. Here, $D$ is the
linear density growth factor. Clearly ${\bf v}_\delta$ contains most
information of RSD cosmology. Nonlinear evolution causes
$W$ at small scales to deviate from the linear theory
prediction. Such deviation is described by the dimensionless 
\be
\tilde{W}(k)\equiv \frac{W(k)}{f(k)}\ .
\ee
This definition is slightly different from that in Paper I. But for
$\Lambda$CDM, in which $f$ is scale
independent, the two definitions are identical. But in
modified gravity models or in models with dark energy fluctuations,
$f$ is scale dependent. $\tilde{W}$ defined in this way, hence, isolates the
impact of nonlinear evolution from the scale dependence in linear
evolution.  We expect $\tilde{W}\leq 1$ in general, even for modified
gravity models and models with dark energy fluctuations. 

We find that $\tilde{W}$ can deviate from unity by $\sim 1-10\%$ at
$k=0.1h/$Mpc. It modifies the Kaiser formula (for dark matter RSD) from
$P_{\delta\delta}(1+f u^2)^2$ to
$P_{\delta\delta}(1+f\tilde{W}u^2)^2$.  So if the $\tilde{W}$ is not
taken into account, $f$ can be biased low by $\sim 10\%$, as predicted
by perturbation theory. A
systematic error of such amplitude is severe for stage IV dark energy
surveys like BigBOSS/MS-DESI \cite{BigBOSSw}, CHIME \cite{CHIMEw},
Euclid \cite{Euclidw} and SKA \cite{SKAw}.
In the present paper we will quantify $\tilde{W}$ more robustly,
through N-body simulations.

${\bf v}_\delta$ causes both large-scale enhancement and small-scale damping in the redshift space clustering. Besides the
$\tilde{W}$ correction, nonlinearities in ${\bf v}_\delta$ induce complicated high
order corrections ($C_G$ and $C_{\rm NG}$) to the Kaiser effect in RSD
modeling (Paper I). These high-order 
corrections involve both the velocity and density fields, so
quantifying them is beyond the scope of this paper and will be left
for future investigation through N-body simulations. In principle,
nonlinearities in ${\bf v}_\delta$ also complicate the FOG
effect.

 ${\bf v}_S$ is also irrotational ($\nabla \times {\bf
    v}_S=0$). But to the opposite of ${\bf v}_\delta$, it is
  uncorrelated with the underlying  density field ($\langle
  \theta_S({\bf x}) \delta({\bf x}+{\bf r})\rangle=0$). This velocity
  component arises from nonlinear evolution. It is the cause of
  density-velocity stochasticity,
\ba
\label{eqn:r}
r_{\delta\theta}(k)&\equiv&
\frac{P_{\delta\theta}(k)}{\sqrt{P_{\delta\delta}(k)P_{\theta\theta}(k)}}=\frac{1}{\sqrt{1+\eta(k)}}\
, \\
\eta(k)&\equiv&
\frac{P_{\theta_S\theta_S}(k)}{P_{\theta_\delta\theta_\delta}(k)}\ . \no
\ea
${\bf v}_S$ causes both  large-scale enhancement and small-scale
damping in the redshift space clustering. But notice that the leading-order large-scale
enhancement has a $u^4$ angular dependence [Eq. (\ref{eqn:RD4})].  It
hence  differs significantly from the Kaiser effect.

${\bf v}_B$ is the curl (rotational) component,
satisfying $\nabla\cdot {\bf v}_B=0$. ${\bf v}_B$ grows only after
orbit crossing occurs. We expect its power to concentrate at small
scales. We also expect that,  to a good approximation, it only damps the
redshift space clustering and only induces the FOG effect.

The present paper aims to quantify the RSD-related statistics of the
three velocity components, through simulations. These statistics
include
\bi
\item the power spectra $\Delta^2_{v_\alpha v_\alpha}\equiv P_{v_\alpha v_\alpha}k^3/(2\pi^2)$,
\item $\tilde{W}(k,z)$ and in particular the
  $\tilde{W}(k,z)$-$\Delta^2_{\delta\delta}(k,z)$ relation,
\item the correlation functions $\psi_{\perp,v_\alpha
    v_\alpha}(r,z)$ and
  $\psi_{\parallel, v_\alpha v_\alpha}(r,z)$,
\item $P(v_{z,\alpha})$, the probability distribution functions (PDFs)
  of ${\bf v}_\alpha$ along the $z$ axis,  and their derived properties such as cumulants and the
  FOG function $D^{\rm FOG}_\alpha$.
\ei
Here,  the subscript $\alpha=\delta, S, B$ denotes the three velocity components.


\section{N-body simulation and velocity assignment method}
\label{sec:simulation}
Given the involved nonlinearities, N-body simulations are needed to
robustly quantify the above  velocity statistics. Furthermore, comprehensive
tests must be designed and performed to
robustly quantify numerical artifacts associated with simulations
themselves and numerical artifacts associated with the velocity
assignment method.

\subsection{N-body simulations}

\begin{table}[b]
\begin{center}
\begin{tabular}{c|c|c|c|cccccc}\hline
Name & $L_{\rm box}/h^{-1}$Mpc & $N_P$ & $m_P/h^{-1}M_\odot$
 & $\Omega_b$ \\\hline
J1200 & 1200 & $1024^3$ & $1.2\times10^{11}$ & 0.045 \\
J300 & 300 & $1024^3$ & $1.9\times10^{9}$ & 0.045 \\
G100 & 100 & $1024^3$ & $6.7\times10^7$  & 0.044 \\\hline
\end{tabular}
\end{center}
\caption{Specifications of the three simulations. Except for the
  slightly different $\Omega_b$, all three have  $\Omega_m=0.268$,
  $\Omega_\Lambda=0.732$, $\sigma_8=0.85$,  $n_s=1$, and
  $h=0.71$. The slight difference in $\Omega_b$ is negligible for the
  velocity statistics presented  in this paper. $N_P$ is the total
  particle number and  $m_P$ is the mass of
  each simulation particle. }
\label{Table:simulation}
\end{table}

From the structure formation theory, we expect that ${\bf v}_\delta$
dominates at linear/large scales while ${\bf v}_S$ and  ${\bf v}_B$
become important at sufficiently nonlinear/small scales. Given the
computation limitation, it would be difficult for a single simulation
to robustly evaluate all three components. A natural remedy is to
combine large box simulations with small box simulations. Large box
simulations help us better understanding the velocity field at large
scales , in particular ${\bf v}_\delta$. Small box simulations have
high mass and force resolution, so  they can probe ${\bf v}_S$
and ${\bf v}_B$ more robustly. Combining these results, we can then
circumvent the computation limitation and  accurately study the
statistical properties of the three  velocity components.

For this reason, we combine three high-resolution (particle number $N_P=1024^3$)
dark matter simulations with box size $L_{\rm box}/h^{-1}{\rm Mpc} =1200$, $300$,
and $100$. For brevity, we will refer to them
 as J1200, J300, and G100, respectively. The simulations utilize standard $\Lambda$CDM
cosmology with flat space and Gaussian initial conditions. Simulation
specifications are listed in Table \ref{Table:simulation}. The adopted
cosmological parameters are identical, expect a slight difference in
the baryon density $\Omega_b$. For the purpose of our work, we can
safely neglect this difference. J1200 and J300 are run
with a particle-particle-particle-mesh (${\rm P^3M}$) code (see
\cite{Jing07} for details). G100 is run
with Gadget2 \cite{Springel01,Springel05}. For J1200 and G100, we
analyze four redshift snapshots, respectively, which are
$z=0,0.501,1.074,1.878,$ of J1200 and $z=0,0.526,1.024,1.947,$ of
G100. Some output redshifts of J1200 differ from corresponding ones of
G100 by $\Delta z/(1+z)\la 3\%$. The linear velocity growth rate is
$fHD$, where $f\equiv d\ln D/d\ln a$ and $D$ is the linear density
growth factor. So these differences in $z$ result in $\sim \frac{1}{2}
\Delta z/(1+z)\la 1.5\%$ \footnote{This is a rough estimation based on
  the approximation $f\sim \Omega_m^{1/2}(z)\propto
\sqrt{\Omega_m(1+z)^3/H^2}$ and $D\sim (1+z)^{-1}$. Under these
approximations,  we have
$fHD\propto (1+z)^{1/2}$. } in velocity.
The major purpose of this paper is to identify the three velocity
components and their basic statistics. For this purpose we can safely
neglect these redshift differences in the current paper. For brevity we
refer to the four snapshots as $z=0.0,0.5,1.0,2.0$ hereafter.
Nevertheless, eventually we need to analyze simulations at
identical redshifts. We also need to run many realizations to robustly quantify
the velocity statistics. For J300, we only analyze its $z=0$ snapshot
since it is mainly used to test our velocity assignment method in
the Appendix (Fig. \ref{fig:velpsall}).

\subsection{The velocity assignment method}
\label{subsec:NPmethod}
It is tricky to properly estimate the volume-weighted
velocity field from velocities of inhomogeneously distributed dark
matter particles or galaxies (e.g. \cite{DTFE96} and references
therein).  In N-body simulations, we only have limited simulation
particles and only have the velocity information at positions of these
particles. However, the velocity where there is no particle is not
necessarily small/negligible. In contrast, since the velocity is determined by the large-scale
matter distribution, velocity in low-density regions/voids can be
large and in general non-negligible. This sampling bias is
difficult to correct from first principles, especially due to the
awkward situation that the velocity and
density fields are neither completely correlated nor completely uncorrelated.

We take a simple procedure to estimate the volume weighted
velocity field in simulations.  For a given grid point, we
assign  the velocity of its nearest dark matter particle/halo/galaxy to it. We call this
method the nearest-particle (NP) method. The probability $P$ for
a particle's velocity to be assigned to a grid is inversely proportional to its ambient
particle number density $n_P$, $P\propto 1/n_P\propto
V$, where $V$ denotes the volume it occupies. So the NP velocity assignment
method indeed constructs a volume-weighted
velocity field  \footnote{{ J. Koda kindly showed us
  their ongoing work \cite{Koda13}, which also proposed the NP method. This work
  also quantified and corrected the alias effect of the NP method.} }.

We emphasize that this velocity
assignment method is different to the widely used nearest-grid-point
(NGP) method in constructing the density field.  For NGP, each
particle looks for its nearest grid point and assigns itself to this grid
point. All particles are used in the assignment. But for NP, each grid
point looks for its nearest particle and is assigned with the particle
velocity. In dense regions, only a fraction of particles are used in
the assignment. In underdense regions, some particles are used
repeatedly and their velocities are assigned to more than one grid
point. In this sense, it does not use all information of
particles. A simple remedy is to shift the grids so another set of
particles is used for the velocity assignment.

 We also emphasize the difference between the NP method and the Voronoi
tessellation (VT) method \cite{DTFE96}.  The first step of the VT
method is to fill the space with polyhedral cells.  Each polyhedron
contains only one particle, and it covers all space points who
consider this particle to be their nearest particle. The velocity
within this polyhedron is assigned to be the velocity of that particle. The
next step is to smooth the velocity field over a given window
function to obtain the velocities on regular grids. We realize that the NP method is
essentially the first step of the VT method. However, unlike the VT
method, the NP method does not apply a smoothing. This is to  avoid artificial
suppression of small-scale random motions, which are {\it real
  signals} of  significant impact on the FOG effect. This artificial suppression of small scale velocity components also
exists in the widely adopted Delaunay tessellation (DT) method
\cite{DTFE96,DTFE09,Pueblas09}. This is the major reason that we do
not adopt this method \footnote{The  Delaunay Tessellation (DT) velocity assignment utilizes a two-step scheme
to get a volume-weighted velocity field:
first it produces a Delaunay tessellation by linking the closest 4
particles to form a tetrahedron network.
The velocity field inside one tetrahedron is determined through
linearly interpolating the velocities at its 4 vertices. Then the velocity field
is smoothed over a given window function to obtain the velocities on regular grids.
The DT method has been proved to be accurate and reliable
at linear and quasi-linear scales (e.g. \cite{DTFE96,Pueblas09,DTFE09}).
However it breaks down at nonlinear regimes where shell-crossings
and  multi-stream flows occur \cite{DTFE09}. In these regions,
the linear interpolation tends to underestimate the velocity by averaging
velocities with different directions. The smoothing procedure
further smoothes
out random (but real) motions at small scales. Since most contribution to
${\bf v}_S$ and ${\bf v}_B$ comes from nonlinear scales, the DT
method could significantly underestimate ${\bf v}_S$ and ${\bf
  v}_B$. }.

The NP method is straightforward to implement. Furthermore, it is
robust in a number of aspects, despite its simplicity: (1) In high-density
regions, NP does not cause underestimation in the (one-dimensional) velocity dispersion
$\sigma_v\equiv \sqrt{\langle v_z^2\rangle}$, since it does not average
over particle velocities. This is important for RSD study, otherwise, 
underestimation of $\sigma_v$ will cause underestimation of FOG.
(2)  It robustly measures ${\bf v}_\delta$, which is of the most
importance for RSD cosmology. For the NP velocity assignment, high-density regions are no problem  because there are always particles
within a grid size. In underdense
regions, the typical distance from a given grid point to its nearest
particle is $(1+\delta)^{-1/3}L_{\rm Box}/N_P^{1/3}$, which can be
large. But in such regions ${\bf
  v}_\delta$ does not vary strongly, since its typical length $L$ of
variation is roughly given by $|v/L|= |\delta W H|$ [Eq. (\ref{eqn:thetam})].  The
requirement to robustly sample ${\bf v}_\delta$ is then
\ba
\frac{L_{\rm Box}}{N_P^{1/3}}&<&\left| \frac{v}{HW}
  \frac{(1+\delta)^{1/3}}{\delta}\right| \\
&=&3{\rm Mpc}/h \ \frac{|v|}{300
  {\rm km}/s}\frac{1}{f}\frac{1}{\tilde{W}} \frac{H_0}{H}\left|\frac{(1+\delta)^{1/3}}{\delta}\right|
\ .\no
\ea
This condition is not difficult to satisfy for typical simulation
specifications today. For example,  for our J1200 simulation, $L_{\rm
  Box}/N_P^{1/3}=1.2$ Mpc$/h$ and the above condition is usually
satisfied. Notice that $\tilde{W}\leq 1$ and $f\leq 1$ ($f\simeq 0.5$ at $z=0$).
An  exception is the very underdense regions with
$\delta\la -0.99$. In these regions, our simulation undersamples the
${\bf v}_{\delta}$ field. But these regions are very rare and we do not
expect this undersampling to be severe for statistically evaluating the
${\bf v}_{\delta}$ field. (3) The NP method does not use the velocity
information of all particles. Nevertheless, the information encoded in
unused particles can be captured by shifting the grids and resampling the
velocity field, when necessary.

\subsection{The velocity decomposition method}
After sampling the velocity field ${\bf v}$ on regular grids by the NP method, we proceed to
decompose ${\bf v}$ into the three eigenmodes. The velocity decomposition
is conveniently operated in Fourier space. First we make the E/B decomposition,
\ba
{\bf v}_E({\bf k})&=&\frac{({\bf k}\cdot{\bf v}({\bf k}))}{k^2}{\bf k},
\\
{\bf v}_B({\bf k})&=&{\bf v}({\bf k})-{\bf v}_E({\bf k})\no \ .
\ea
Then we decompose ${\bf v}_E$ into ${\bf v}_\delta$ and ${\bf v}_S$ [
Eq. (\ref{eqn:thetam})] \footnote{The sign of ${\bf v}_\delta$ expression being plus or minus depends on the definition of Fourier transform. Our definition for Fourier transform is $\delta({\bf k})\equiv\int \delta({\bf x})\exp(i{\bf k}\cdot {\bf x})d^3x$.},
\ba
{\bf v}_\delta({\bf k})&=&-i\frac{H(z)\delta({\bf k}) W(k)}{k^2}{\bf k}\
,\\
{\bf v}_S({\bf k})&=&{\bf v}_E({\bf k})-{\bf v}_\delta({\bf k})\ .\no
\ea
Here $H(z)$ is the Hubble parameter at redshift $z$, and the window function $W$ is calculated in advance by $W=P_{\delta\theta}/P_{\delta\delta}$.
The density field $\delta$ is sampled by the NGP method on the same regular grids as the
velocity field. When calculating $P_{\delta\delta}$, we do not correct the shot
noise term since it is negligible for our high particle number density simulation.
Also we do not correct other numerical artifacts like smoothing and
alias effects, which are subdominant to systematic errors induced by
the velocity assignment method. Hence, numerical artifacts quantified by our designed tests (Appendix
\ref{sec:NPtest}) receive minor contribution from those in the density field.

\subsection{Testing the NP method}
In Appendix \ref{sec:NPtest} we design and carry out several tests to
verify the robustness of the NP method for our study.

 First we  compare the power spectra of the three velocity components between different
box sizes, grid numbers, and particle numbers. Discrepancies there
diagnose sampling bias. In particular, the NP method  becomes exact in the limit $N_P/L_{\rm box}^{3}\rightarrow \infty$. By constructing the velocity field using a fraction of
particles and observing its dependence on the particle numbers, we can
evaluate its accuracy. If the results converge, we then have
confidence of vanishing sampling bias.   Detailed comparison is given in Appendix \ref{sec:NPtest}.

In brief,  we find that NP method can construct ${\bf
v}_\delta$ and ${\bf v}_S$ fields quite reliably. However, for large box simulations
(e.g. J1200),  the ${\bf v}_B$ amplitude is significantly
overestimated at all scales due to a shot-noise-like alias effect induced by
sparse sampling \cite{Pueblas09}. However, tests on small box simulation G100 show reasonable
convergence of ${\bf v}_B$. So the G100 simulation provides reasonably
accurate measure of ${\bf v}_B$. Hence, in combination of J1200 and G100, we have reliable measures of all three velocity
components.

We also test  the NP method against the fiducial velocity
field of known statistics as input (\S
\ref{subsec:rigoroustest}). This test is particularly good at
highlighting leakage between the three velocity components. We do find
such leakages. We quantify their amplitudes as a function of scales
and identify regions where these leakages are under control.

These tests show that, combining J1200 and G100, we can control
numerical artifacts and reliably measure all three
velocity components. Even better, the impacts of these numerical
artifacts on RSD cosmology are further
reduced for a number of reasons. First, the majority of cosmological information resides in
relatively large scales and especially in ${\bf v}_\delta$. This
component is accurately measured. Second, although accurate
measurement of ${\bf v}_B$ is more challenging,  its impact on
RSD is fully captured by  its velocity dispersion $\sigma_{v_B}$ (Sec.
\ref{sec:results}), which can be  treated as a free
parameter to be fitted against data. Even simpler, Sec. \ref{sec:results}
will show that setting $\sigma_{v_B}=0$  is already sufficiently
accurate for RSD modeling.

\bfi{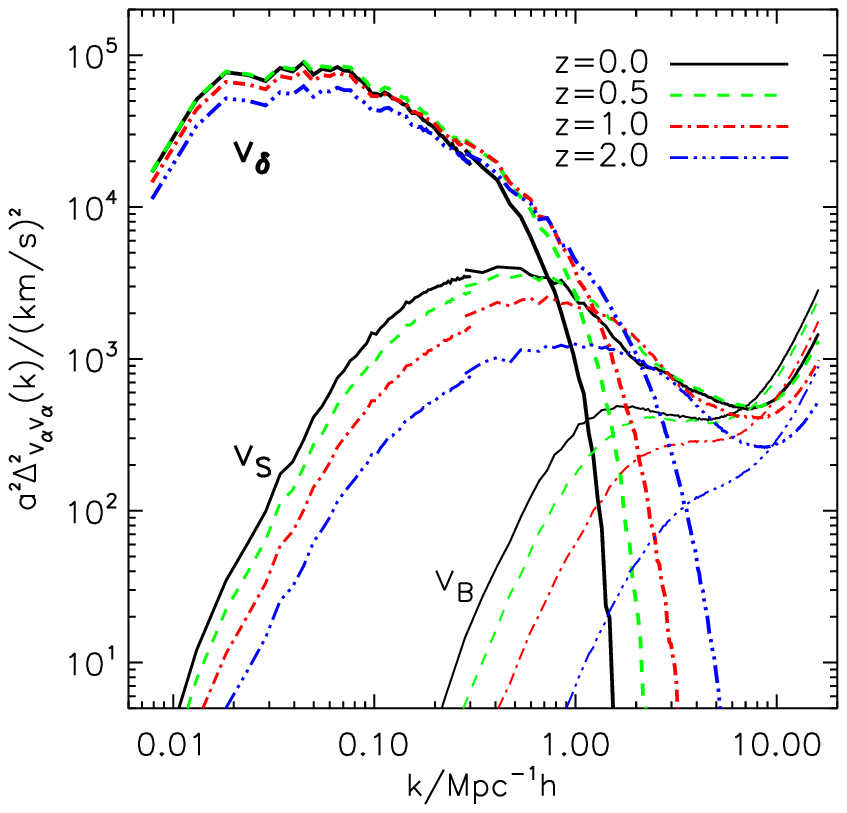}
\caption{The proper velocity power spectrum $a^2\Delta^2_{v_\alpha
    v_\alpha}(k,z)$ ($\alpha=\delta,S,B$) at $z=0,0.5,1.0,2.0$. As a note, all symbols of ${\bf v}$  in
   Paper I and this paper refer to the {\it comoving} peculiar velocity. But in order to
   better show the redshift evolution, we plot the physical
   velocity power spectra $a^2\Delta^2_{v_\alpha v_\alpha}$ instead.
The thick,
  intermediate thick, and thin lines represent $a^2\Delta^2_{v_\delta v_\delta}(k,z)$,
  $a^2\Delta^2_{v_S v_S}(k,z)$, and $a^2\Delta^2_{v_B v_B}(k,z)$, respectively.
  We combine the J1200 simulation and the G100
  simulation to reduce numerical artifacts. $a^2\Delta^2_{v_\delta v_\delta}$ and
  $a^2\Delta^2_{v_Sv_S}$ at
  $k<0.3h/{\rm Mpc}$  are from J1200, while those of $k>0.3h/{\rm Mpc}$ are from G100.
   $a^2\Delta^2_{v_Bv_B}$ is from G100.
   As shown in this figure, $a^2\Delta^2_{v_\delta v_\delta}$ at $z=0$ and $z=0.5$ overlap
   largely. This is because $a{\bf v}_\delta$ ceases to grow after
   $z\simeq0.2$ as the cosmological constant slows down the structure
   growth. Unless otherwise specified,  the measurement is done on
   $N_{\rm grid}=512^3$ grid.}
\label{fig:velps}
\efi

\bfi{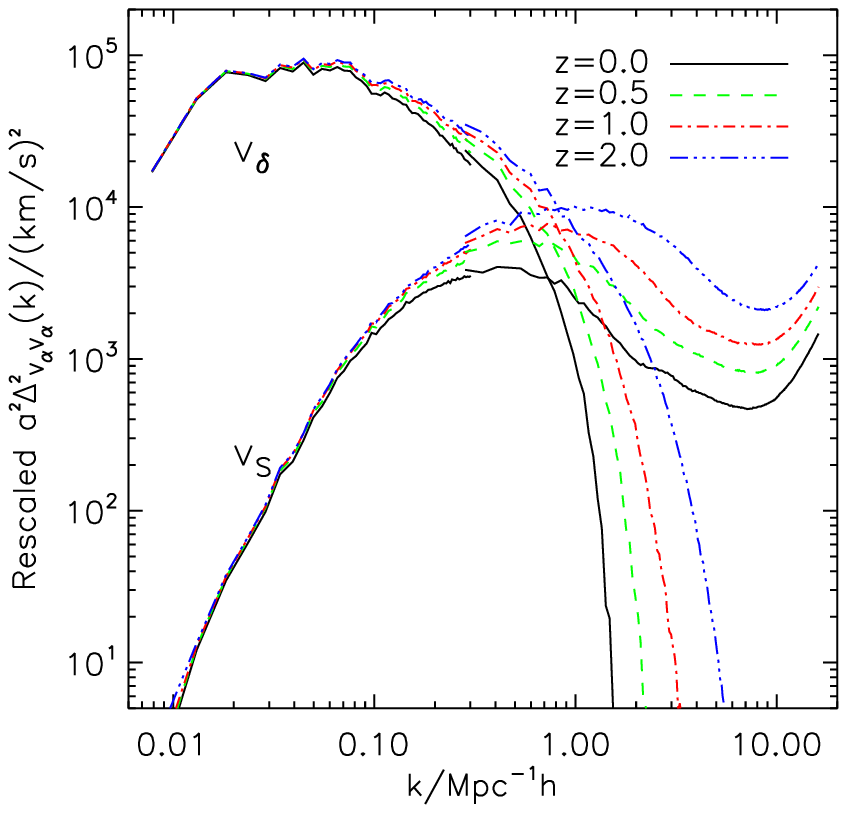}
\caption{
  Rescaled $a^2\Delta^2_{v_{\delta,S}
    v_{\delta,S}}$ to demonstrate their redshift evolution. (1) $\Delta^2_{v_\delta v_\delta}(z)$
  is rescaled by the linear velocity growth factor
  $D^2_{v_\delta}(z=0)/D^2_{v_\delta}(z)$ [Eq. (\ref{eqn:Dvz})].
  Nonlinear evolution is
  visible at $k\ga 0.03 h/$Mpc and becomes non-negligible at $k\ga
  0.1h/$Mpc. (2)
  $\Delta^2_{v_Sv_S}(z)$ is rescaled by the third-order perturbation
  prediction $(D^2fH)^2(z=0)/(D^2fH)^2(z)$ [Eq. (\ref{eqn:vsz})]. }
\label{fig:velpsd}
\efi

\section{Statistics of ${\bf v}_{\delta,S,B}$}
\label{sec:results}
Understanding the statistical properties of the three velocity
components facilitates the RSD modeling and 
velocity reconstruction (Paper I). For this purpose, we measure the one-point and two-point
velocity statistics from simulations. (1) The velocity
power spectra $\Delta^2_{v_\alpha v_\alpha}(k,z)$ ($\alpha=\delta, S,
B$) go directly into the redshift space matter power
spectrum.  They also quantify the scale dependence and
redshift evolution of these velocity components.  They are useful to
understand the physical  origins of these velocity components. The
window function $W$ is a derived quantity of these power spectra. (2) Velocity correlation functions
$\psi_{\parallel,v_\alpha v_\alpha}$ and $\psi_{\perp,v_\alpha
  v_\alpha}$ ($\alpha=\delta, S, B$)
quantify the velocity correlation length, $L_\alpha$. These correlation lengths are crucial in judging at
which separation we can treat the velocities at two positions as
independent. Basically, if the scale of interest $k\ll 2\pi/L_\alpha$,
we can safely ignore the intrinsic clustering of the ${\bf v}_\alpha$
field. In this limit, we can treat ${\bf v}_\alpha$ as a random field in RSD
modeling, whose impact is fully
captured by a damping function $D^{\rm FOG}_\alpha$. Otherwise, we
have to take their intrinsic clustering into account, which
contributes extra power to the redshift space clustering. (3) Finally, we will
quantify the non-Gaussianities of the velocity fields, through the
one-point PDF and the reduced cumulants $K_j$. These
non-Gaussianity measures are highly relevant for  modeling $D^{\rm
  FOG}_\alpha$, which describes the FOG effect.

\subsection{The velocity power spectrum}
\label{subsec:ps}

\bfi{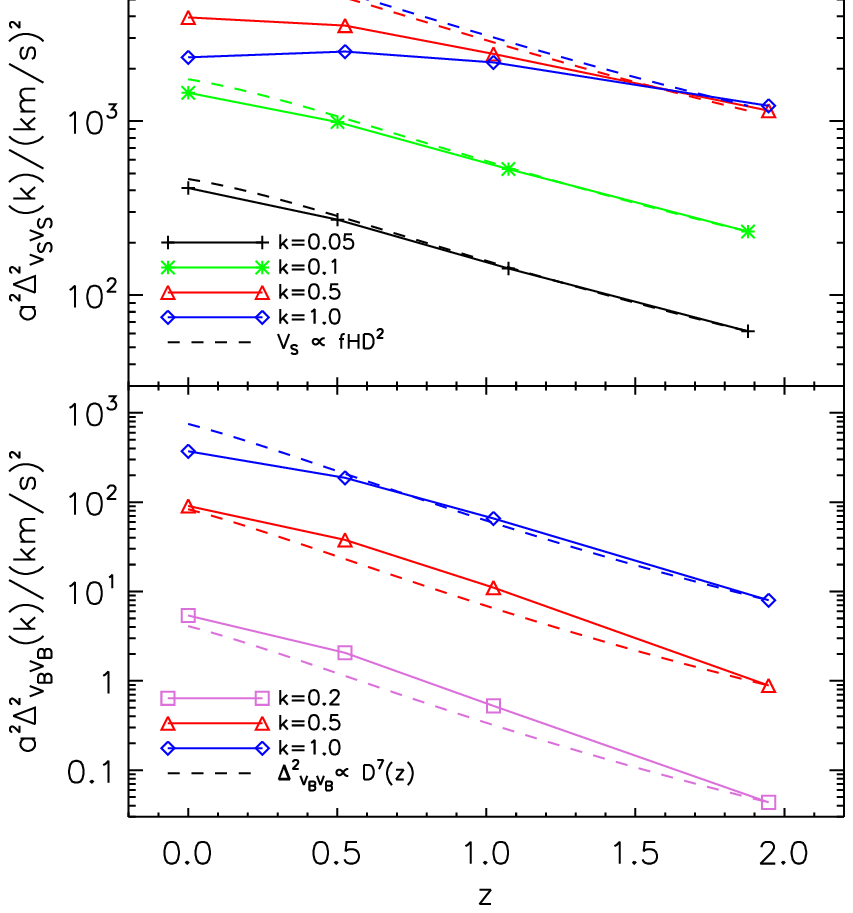}
\caption{Redshift evolution in $\Delta^2_{v_\alpha v_\alpha}$ ($\alpha=\delta,S,B$).
 Dashed lines correspond to predictions of the linear perturbation
 theory (${\bf v}_\delta\propto fHD$), third-order perturbation theory (${\bf
   v}_S\propto fHD^2$), and the finding by \cite{Pueblas09} on ${\bf
   v}_B$ ($\Delta^2_{v_Bv_B}\propto D^7$). }
\label{fig:Dz}
\efi

\bfi{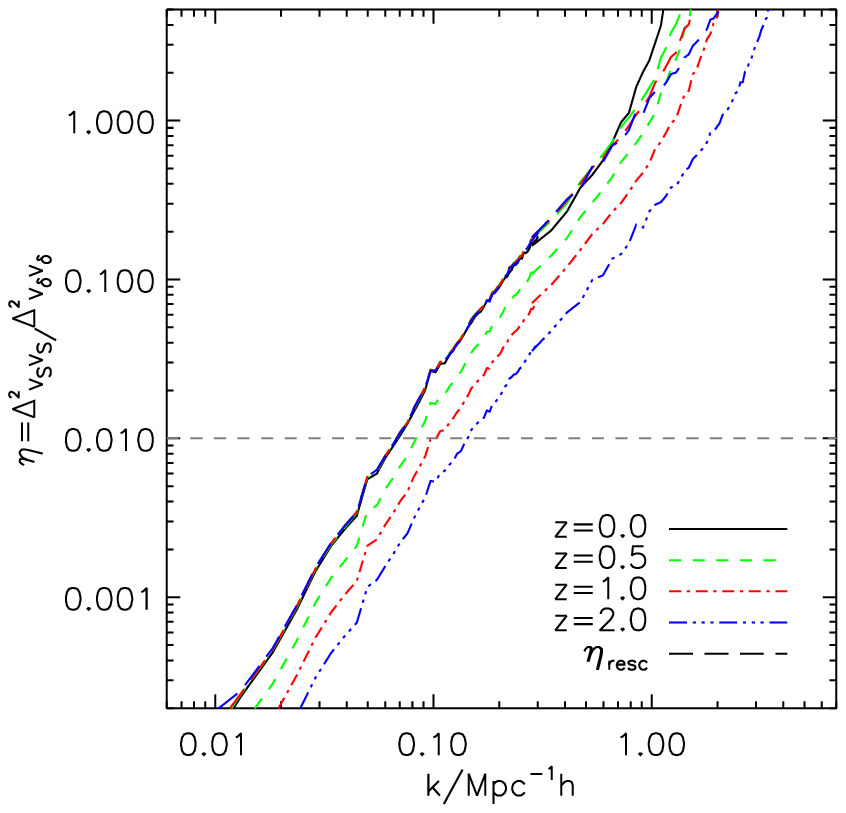}
\caption{$\eta\equiv P_{v_Sv_S}/P_{v_\delta v_\delta}$ is shown at $z=0,0.5,1.0,2.0$. $\eta$
  quantifies the velocity-density stochasticity ($r_{\delta\theta}=1/\sqrt{1+\eta}$). It also shows the
  relative importance of ${\bf v}_S$ with respect to ${\bf
    v}_\delta$. Future surveys require $1\%$ accuracy in RSD modeling
  and hence require the inclusion of ${\bf v}_S$ at $k\ga
  0.1h/$Mpc.  To compare with the perturbation theory prediction
  ($\eta\propto D^2$), we plot $\eta_{resc}=\eta(z)\times
  D^2(z=0)/D^2(z)$ (long dashed lines). The prediction works well such
  that the rescaled lines largely overlap with each.  We find that, to a good
  approximation, $\eta(k,z)\propto D^2(z)k^{n_\eta}$, with
  $n_\eta\simeq 2.2$ at $k\la 0.7h/$Mpc. }
\label{fig:eta}
\efi

The velocity power spectra at $z=0.0, 0.5, 1.0, 2.0$ are shown in Fig.
\ref{fig:velps}. These results confirm our speculation in Paper I, based upon the
structure formation theory.  It shows that ${\bf v}_\delta$ is the
dominant component at linear and mildly nonlinear
scales. Perturbation theory predicts that it evolves linearly at sufficiently
large scales, with the linear velocity growth factor
\be
\label{eqn:Dvz}
D_{v_\delta}(z)=f(z)H(z)D(z)\ .
\ee
 Figures. \ref{fig:velpsd} and \ref{fig:Dz} verify
this linear evolution at  $k<0.1 h/$Mpc. However, linear
perturbation theory quickly loses its predicting power at $k> 0.1
h/$Mpc. Even at $k=0.1h/$Mpc, impact of nonlinear evolution is visible.

\bfig{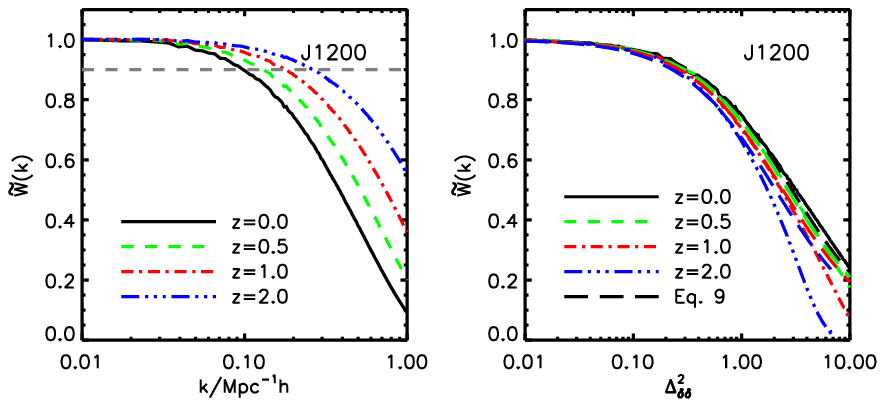}
\caption{
  \textit{Left panel}: the $\tilde{W}(k,z)-k$ relation, calculated
  from the J1200 simulation. $\tilde{W}$ changes from unity at $k\rightarrow 0$
   to zero at $k\rightarrow \infty$. $1-\tilde{W}\ga 10\%$
   at $k=0.1h$/Mpc and $z\alt 0.5$, indicating a significant systematic error in
   RSD cosmology even for stage III dark energy surveys like BOSS and eBOSS.
   \textit{Right panel}: the $\tilde{W}(k,z)-\Delta^2_{\delta\delta}$
   relation. Long dashed lines represent the fitting
   formula [Eq. (\ref{eqn:Wfit})] with the best-fit
   $\Delta_\alpha$ listed in Table. \ref{Table:fitting}. Comparing to the $\tilde{W}$-$k$
   curves in the left panel, the redshift dependence of $\tilde{W}$-$\Delta^2_{\delta\delta}$
   curves is greatly reduced.}
\label{fig:wk}
\efig

Nonlinear evolution in the velocity and density fields drive
$\tilde{W}\equiv W/f$ to decrease from unity. $\tilde{W}$ is one of the
most important properties to describe the velocity field, to model
RSD,  and to reconstruct velocity in spectroscopic surveys. Hence we devote the
whole Sec. \ref{subsec:W} to discuss it.

Nonlinear evolution also induces stochasticity in the velocity-density
relation and causes the emergence of ${\bf v}_S$
(Fig. \ref{fig:velps}).  It eventually dominates over ${\bf v}_\delta$
at 
$k=1h/$Mpc and $z=0$. To better show its impact, Fig. \ref{fig:eta} plots the ratio
$\eta\equiv P_{v_S v_S}/P_{v_\delta v_\delta}$.  Stage IV dark energy projects such as
BigBOSS/MS-DESI, CHIME, Euclid, and SKA can achieve $1\%$ level statistical precision for
the velocity measurement through RSD. So once $\eta>1\%$, the ${\bf
  v}_S$ component becomes non-negligible in RSD modeling.  $\eta$ reaches $\simeq 1\%$
at $k=0.1h/$ Mpc and $z\alt 1$ (Fig. \ref{fig:eta}).  Even at high
redshift $z=2$, $\eta\simeq 1\%$ at $k=0.2h/$ Mpc. These
results agree with calculation by high-order perturbation theory
(Fig. 1, Paper I). It confirms our conclusion in Paper I that, in general ${\bf v}_S$
is a non-negligible velocity component, even at scales which are often
considered as linear. Its contribution to the redshift space matter power spectrum
will be quantitatively studied in future works. Future work will also explore information
encoded in ${\bf v}_S$. For example, it may be used to probe the
environmental dependence of modified gravity theories (Paper I).

Perturbation theory is also useful to understand the redshift evolution of
${\bf v}_S$. From the continuity equation,
\be
\dot{\delta}+\nabla \cdot (1+\delta){\bf v}=0\ ,
\ee
the leading-order contribution to ${\bf v}_S$ comes from $\dot{\delta}^{(2)}$ and
$\delta^{(1)}{\bf v}^{(1)}$. Here,  $\delta=\sum_i \delta^{(i)}$ is the sum over contributions of $i$th
order density component. To a good approximation, $\delta^{(i)}\propto
D^i$ \cite{Bernardeau02}. So both
contributions to ${\bf v}_S$ evolve as $\propto D^2 fH$.  Unlike
the case of the density power spectrum, the leading-order contribution
to $P_{v_S v_S}$ do not have contribution from third-order components
(e.g., $\langle \dot{\delta}^{(3)}v^{(1)}\rangle$) because
  ${\bf v}_S$ vanishes at linear order (${\bf v}^{(1)}_S=0$). So the perturbation theory
  predicts that
\be
\label{eqn:vsz}
P_{v_S v_S}(z)\propto (D^2fH)^2\ .
\ee
Figures \ref{fig:velpsd} and \ref{fig:Dz} verify this relation at  $k\la
0.1 h/$Mpc. But it quickly loses accuracy toward higher $k$ (smaller
scales).

Surprisingly, perturbation theory works much better to understand
$\eta\equiv P_{v_Sv_S}/P_{v_\delta v_\delta}$.  It predicts
$\eta\propto D^2$ [Eq. (\ref{eqn:Dvz}) and
\ref{eqn:vsz})]. Figure \ref{fig:eta} shows that it works even at
$k=0.7h/$Mpc. Another interesting finding to report is the
surprisingly simple scale dependence of $\eta(k,z)$, despite
complexities  in shapes of both ${\bf v}_\delta$ and ${\bf v}_S$
(Fig. \ref{fig:velps}). It is well described by a power law.  Over the
range $k\in (0.01, 1)h/$Mpc,
$\eta(k)\propto k^{n_\eta}$ with $n_\eta\simeq 2.2$ (Fig. \ref{fig:eta}).  Whether
these behaviors  are coincident or generic
requires further investigation. If these behaviors are generic, they
can be utilized to further reduce degrees of freedom in RSD modeling.

${\bf v}_B$, the curl component, grows only where shell crossing and
multistreaming happen. This is the place where perturbation theory,
which is based on the single fluid approximation,  breaks down.
So we lose a powerful tool to understand its behavior. Nevertheless,
\cite{Pueblas09} found that  $P_{v_Bv_B}(z)\propto D^7(z)$. Our
results confirm this relation 
(Fig. \ref{fig:Dz}).  Figure \ref{fig:velps} shows that ${\bf
v}_B$ grows later than ${\bf v}_S$. It is less than $1\%$ of ${\bf
v}_\delta$ at $k=0.3h/$Mpc. It is subdominant to ${\bf v}_S$ at $k\la
3h/$Mpc. Our convergence tests presented in the Appendix do not find
significant numerical artifacts on ${\bf v}_B$ measured from the $100
$Mpc$/h$ G100 simulation at $k\la 3h/$Mpc. So the above results should
be reliable. We may also expect that  the velocity field becomes
completely randomized at sufficiently small scales, so
$P_{v_Bv_B}\rightarrow 2 P_{v_Sv_S}$. We do find this sign of
equipartition at $k\sim 10h/$Mpc. However, numerical artifacts at
these regimes are non-negligible, as shown in the
Appendix. Simulations with resolution higher than G100 are required to
study this issue.


\subsection{The window function $\tilde{W}(k,z)$}
\label{subsec:W}


\bfig{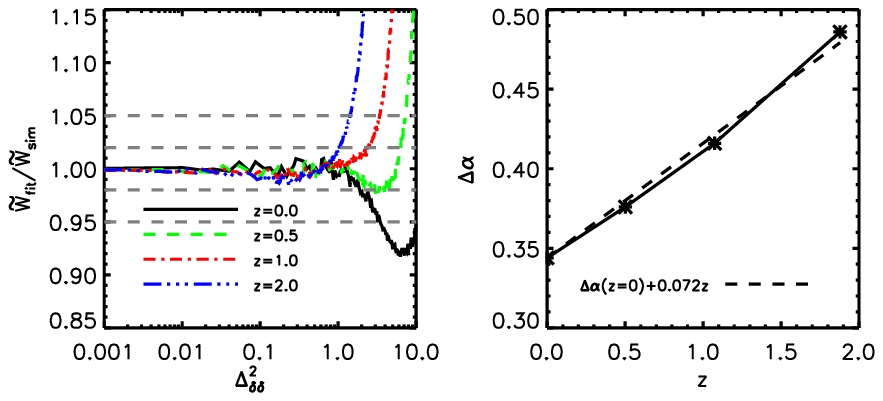}
\caption{
  \textit{Left panel}: $\tilde{W}(k)_{\rm fit}$/$\tilde{W}(k)_{\rm sim}$ with best
  fitted $\Delta_\alpha$ at each redshift. The fitting formula (Eq. \ref{eqn:Wfit})
  achieves an accuracy within $2\%$ at scales of interest for RSD cosmology
  ($\Delta^2_{\delta\delta}\leq1$) at all redshifts.
  \textit{Right panel} shows  the best-fit $\Delta \alpha$ at four
  redshifts. The redshift dependence in $\Delta \alpha$ is
  essentially linear relation,  with a slope $b\simeq0.072$ (dashed line).}
\label{fig:fit}
\efig

The window function $W(k)$ [Eq. (\ref{eqn:thetam})] and the normalized one,
$\tilde{W}(k)\equiv W(k)/W(k\rightarrow 0)=W(k)/f$, are of crucial
importance for the following reasons in understanding and reconstructing the velocity field and
in improving RSD modeling (Paper I). (1) $\tilde{W}$ describes the impact of the
nonlinear evolution on the velocity-density relation. (2)  $\tilde{W}$
quantifies a major systematic error in RSD cosmology. We have shown
that the leading term in the redshift space matter power spectrum $P^s_{\delta\delta}(k,u)$
is $P_{\delta\delta}(1+f\tilde{W}(k)u^2)^2$. Hence the widely
adopted  Kaiser formula underestimates $f$ by a factor
$\tilde{W}(k)\leq 1$. Perturbation theory predicts a $\sim 10\%$ bias
at $k=0.1h/$Mpc and $z=0$ (Paper I), much larger than the statistical error
associated with stage IV dark energy surveys. (3) $W$
behaves as a window function exerting on the density field to
reveal the underlying velocity field. This deterministic
function can be inferred from RSD in spectroscopic redshift surveys in
less model dependent way. So it is essential in three-dimensional peculiar velocity
reconstruction in redshift surveys, {\it at cosmological distances}.

Figure \ref{fig:wk} shows $\tilde{W}(k)$ at different redshifts measured
in the J1200 simulation.
It confirms our theoretical prediction using third-order Eulerian
perturbation theory (Paper I), which is applicable at $k\la 0.2 h/$Mpc. As
expected, $\tilde{W}$ changes from unity at $k\rightarrow 0$ to zero
at $k\rightarrow \infty$. As explained in Paper I, $1-\tilde{W}$
quantifies a systematic error in $f$. Since stage IV
dark energy surveys have the potential to measure $f$ to $\sim 1\%$
level accuracy, this $\tilde{W}$-induced systematic error becomes
significant, even at relatively high redshift $z=2$ and pretty linear
scale $k=0.1h/$Mpc.  The situation worsens towards lower redshifts and
smaller scales. For example, $1-\tilde{W}\ga 10\%$ at $z\alt 0.5$. It is
already significant for stage III dark energy surveys like BOSS and
eBOSS \cite{SDSS1,SDSS2,SDSS3}.
This systematic error may contribute a significant fraction
to the tension in $f$ between existing measurements and the
prediction from Planck cosmology \cite{Macaulay13}.

\begin{table}[b]
\begin{center}
\begin{tabular}{c|cccccc}\hline
Redshift & $\Delta \alpha$ & $\chi^2$ & $\chi^2_{dof}$ & $N_{\rm data}$\\\hline

$z=0$ & 0.344 & 1.001 & 0.026 & 38\\
$z=0.5$ & 0.376 & 1.0 & 0.019 & 54\\
$z=1$ & 0.416 & 1.001 & 0.012& 82 \\
$z=2$ & 0.486 & 1.01 & 0.007 & 145\\\hline
\end{tabular}
\end{center}
\caption{Fitting parameters. $\chi^2=\sum (\tilde{W}_{\rm
    fit}-\tilde{W}_{\rm sim})^2$.}\label{Table:fitting}
\end{table}

We expect the degrees of freedom in $\tilde{W}$ is limited. The
perturbation theory predicts \cite{Bernardeau02}, for a power law initial power spectrum
with power index $n$,
\ba
\tilde{W}(k,z)&=& \frac{1+\alpha_{\delta\theta}(n)\Delta^2_{L}(k,z)+O(\Delta^4_L)}{1+\alpha_{\delta\delta}(n)\Delta^2_{L}(k,z)+O(\Delta^4_L)}\\
&\simeq& \frac{1}{1+\Delta \alpha(n)\Delta^2_L(k,z)}\no \ .
\ea
Here, $\Delta^2_L$ is the linear matter power spectrum variance and
$\Delta^2_{\rm NL}$ is the nonlinear one. The $\alpha$ symbols follow
the notation in \cite{Bernardeau02} and $\Delta \alpha(n)\equiv
\alpha_{\delta\delta}(n)-\alpha_{\delta\theta}(n)$. However,  in reality, the power index
depends on $k$. So $\Delta \alpha$ is a function of both $k$ and
$z$. In this case $n$ is often approximated as the
effective power index at the nonlinear scale $k_{\rm NL}$, defined through $\Delta_L^2(k_{\rm NL},z)=1$. Even so, $\Delta \alpha=\Delta \alpha(n_{\rm
  eff}(k_{\rm NL}(z)))$, a function of redshift. On the other hand,
the neglected terms of $O(\Delta^4_L)$ imply a stronger dependence
than $\Delta^2_L$.  These considerations motivate us to
propose the following fitting formula:
\ba
\label{eqn:Wfit}
\tilde{W}(k,z)&=&\frac{1}{1+\Delta \alpha(z)\Delta^2_{\rm
    NL}(k,z)}\\
&\equiv& \frac{1}{1+\Delta \alpha(z)\Delta^2_{\delta\delta}(k,z)}\no \ .
\ea
Notice that we have replaced the linear matter power spectrum with the
nonlinear one. Although we still adopt the symbol $\Delta \alpha$, it
is no longer a prediction from perturbation theory. Instead, it shall
be treated as a free function to be fitted against simulation or
observation. Nevertheless, from the above argument, we do not expect a
strong redshift dependence in $\Delta \alpha$.

To check for the above arguments, we plot $\tilde{W}$ against
$\Delta^2_{\delta\delta}$ (right panel, Fig. \ref{fig:wk}). Comparing to the $\tilde{W}$-$k$ curves in the left panel,
we find that the redshift dependence of $\tilde{W}$-$\Delta^2_{\delta\delta}$ curves is greatly
reduced. Curves of different redshifts almost overlap with each other at
$\Delta^2_{\delta\delta}\la 1$. This behavior suggests that $\tilde{W}$
is mainly determined by $\Delta^2_{\delta\delta}$.  Impacts of any
other factors should be minor.
\begin{figure*}[t]
\epsfxsize=15cm
\epsffile{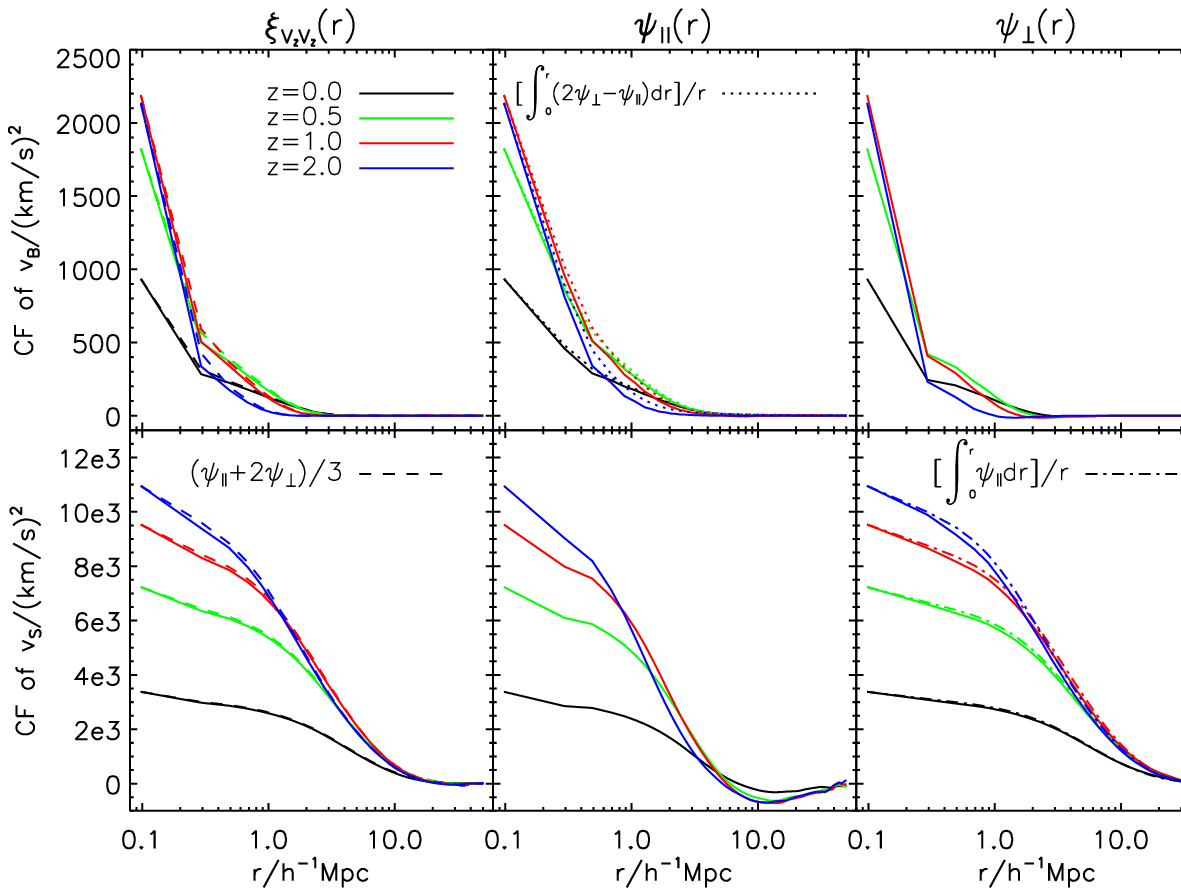}
\end{figure*}

\begin{figure*}[t]
\epsfxsize=15cm
\epsffile{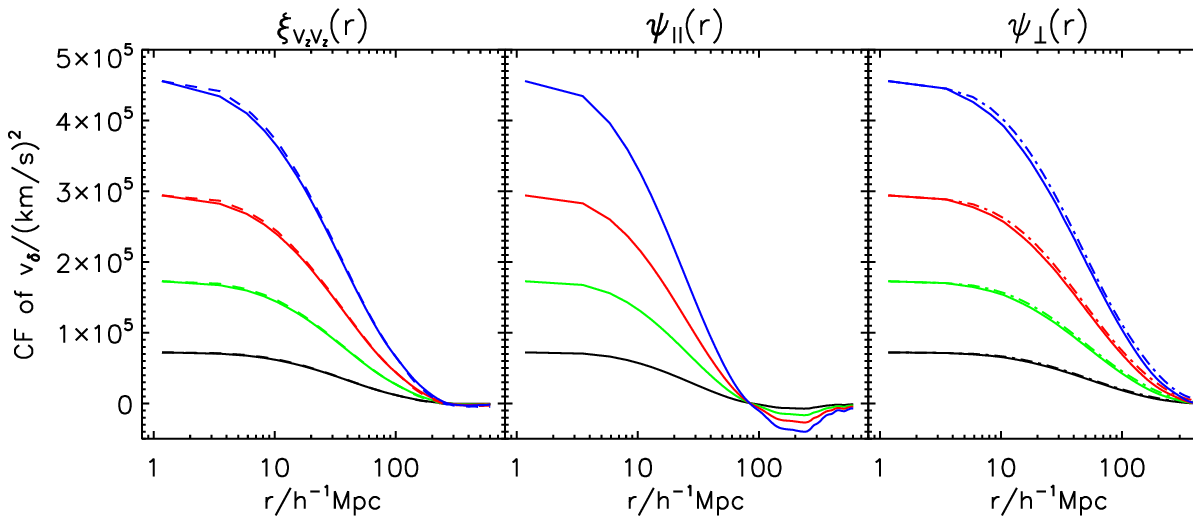}
\caption{Three types of correlation function (CF) are shown by solid lines in different
  columns. Different line colors represent different redshifts.
  CFs of ${\bf v}_\delta$ are calculated from the 
  J1200 simulation. CFs of ${\bf v}_{S,B}$ are calculated from the G100 simulation.
  They are calculated by Fourier transforming the velocity power spectra.
  ${\bf v}_\alpha$ has the largest $O(100)$Mpc correlation length. ${\bf v}_S$ has a
  correlation length of $O(10)$Mpc. ${\bf v}_B$ has the smallest $O(1)$Mpc
  correlation length. The data point on the left end of each line is actually
  $\xi_{v_{z,\alpha}v_{z,\alpha}}(r=0)=\sigma^2_{v_\alpha}$.
  \textit{Left panel}: $\xi_{v_zv_z}(r)\equiv \langle \xi_{v_zv_z}
  ({\bf r})\rangle_{\bf r}$. The dashed lines are calculated by Eq. (\ref{eqn:xizz}) and
  verify this relation.
  \textit{Middle panel}: $\psi_{\parallel}(r)$ is the correlation function when
  both velocities are  along ${\bf r}$. The dotted lines verify Eq. (\ref{eqn:psib}).
  \textit{Right panel}: $\psi_{\perp}(r)$ is the correlation function when both velocities
  are perpendicular to ${\bf r}$. The dot-dashed lines verify Eq. (\ref{eqn:psids}).}
\label{fig:cf}
\end{figure*}

We then obtain the best-fit value of $\Delta \alpha$ at each
redshift. Table \ref{Table:fitting} lists the best-fit $\Delta
\alpha$, 
the associated $\chi^2$ and reduced $\chi^2$. We limit the fitting to
$\Delta^2_{\delta\delta}\leq 1$, where robust cosmology based on RSD
is promising. Furthermore, the proposed fitting formula is not
expected to work well at $\Delta^2_{\delta\delta}\gg 1$ (Fig. \ref{fig:fit}).

It turns out that the proposed fitting formula works excellently at
$\Delta^2_{\delta\delta}<1$ (Fig. \ref{fig:fit}). It achieves an
accuracy within $2\%$ at scales of interest for RSD cosmology
($\Delta^2_{\delta\delta}\leq1$).

The best-fit $\Delta \alpha$ varies weakly with redshift. The redshift
dependence can be excellently approximated as linear,
\be
\label{eqn:Delta}
\Delta \alpha(z)\simeq \Delta \alpha(z=0)+bz\ ,
\ee
with $b\simeq 0.072$ (Fig. \ref{fig:fit}).

By far we have illustrated the possibility of finding a simple fitting
formula to accurately model $\tilde{W}(k,z)$. We caution that, although the proposed
form [Eq. (\ref{eqn:Wfit})] works very well, it may not necessarily be the best-fitting
formula. What we
really want to demonstrate here is that  $\tilde{W}$ has very
limited degrees of freedom such that one or two fitting parameters are
sufficient to model it to high accuracy. This means that we can
efficiently reduce the $\tilde{W}$-induced systematic error without
significantly inflating the statistical error in $f$. Whether we
can find  a more physically
motivated  and hence more generic and more accurate fitting formula is
an issue for further investigation.

\subsection{The velocity correlation function}
\label{subsec:cf}

RSD modeling in Paper I requires information on the velocity
correlation function. The correlation function $\xi_{ij}({\bf
  r})\equiv \langle v_i({\bf x}_1)
v_j({\bf x}_2)\rangle$
between the $i$th velocity component at position ${\bf x}_1$ and
$j$th velocity component at ${\bf x}_2={\bf x}_1+{\bf r}$ can be decomposed into two
correlation functions $\psi_{\perp}$ and $\psi_{\parallel}$ \cite{Peebles80},
\be
\label{eqn:cf}
\xi_{ij}({\bf r})=\psi_{\perp}(r)\delta_{ij}+\left[\psi_{\parallel}(r)-\psi_{\perp}(r)\right]\frac{r_ir_j}{r^2}
\ .
\ee
Here $\psi_{\parallel}$ is the correlation function of the velocity
components along ${\bf
  r}$ and $\psi_{\perp}$ is the one of velocity components
perpendicular to  ${\bf r}$. $i=x,y,z$ denote
the Cartesian axis. ${\bf r}\equiv {\bf x}_1-{\bf x}_2$ is
the pair separation vector.

$\psi_{\parallel}$ and $\psi_{\perp}$ do not depend on the choice of
coordinate system. The two are not
independent. For a potential flow (like ${\bf
  v}_\delta$ and ${\bf v}_S$), we have the textbook result  \cite{Peebles80},
\ba
\label{eqn:psids}
\psi_{\parallel}(r)&=&\frac{d(r\psi_{\perp}(r))}{dr}\ ,\\
\psi_{\perp}(r)&=&H^2\int \Delta^2_{\theta\theta}\left[\frac{\sin(kr)}{(kr)^3}-\frac{\cos(kr)}{(kr)^2}\right]\frac{dk}{k^3}\ .\no
\ea
Paper I derives the relation for a curl velocity field like ${\bf v}_B$,
\ba
\label{eqn:psib}
\psi_{\perp}(r)&=&\psi_{\parallel}(r)+\frac{1}{2}r\frac{d\psi_{\parallel}(r)}{dr}\ ,\\
\psi_{\parallel}(r)&=&\int
\Delta^2_{v_Bv_B}\left[\frac{\sin
    (kr)}{(kr)^3}-\frac{\cos (kr)}{(kr)^2}\right]\frac{dk}{k}\ .\no
\ea

What is relevant for RSD modeling is $\xi_{v_zv_z}({\bf r})$, assuming
the line of sight as the $z$ axis. One can easily verify that
\ba
\xi_{v_zv_z}({\bf r}=(0,0,r))&=&\psi_\parallel(r)\ ,\\
\xi_{v_zv_z}({\bf r}={\bf r}_\perp)&=&\psi_\perp(r)\ . \no
\ea
One can also prove that
\ba
\label{eqn:xizz}
\xi_{v_zv_z}(r)\equiv \langle \xi_{v_zv_z}({\bf r})\rangle_{\bf
  r}=\frac{1}{3}\left(\psi_\parallel(r)+2\psi_\perp(r)\right)\ .
\ea
Figure \ref{fig:cf} shows the simulated $\xi_{v_zv_z}(r)$ (left column),
$\psi_{\parallel}(r)$ (middle column) and  $\psi_{\perp}(r)$ (right column).
It also verifies the above
relation (dashed lines in left column).
Equations (\ref{eqn:psids}) and (\ref{eqn:psib}) are also verified against
their integral forms [Eq. (\ref{eqn:psids}) by dot-dashed lines in right column,
Eq. (\ref{eqn:psib}) by dotted lines in middle column]:
\ba
r\psi_{\parallel,B}&=&\int_0^r(2\psi_{\perp,B}-\psi_{\parallel,B})dr \ ,\no\\
r\psi_{\perp,S,\delta}&=&\int_0^r\psi_{\parallel,S,\delta}dr \no\  .
\ea
There are slight deviations in the comparison, likely
caused by the simplest trapezoidal rule for integration, or sparse
sampling in low-$k$ region. 

Figure \ref{fig:cf} verifies our speculation in Paper I on the
correlation lengths of the three velocity components. We see that ${\bf v}_\delta$ has the largest
correlation length of $O(100)$ Mpc, in
concordance with the fact that its power spectrum peaks at $k\sim
0.05h/$Mpc. Due to this large correlation length and due to its
complete correlation with the  density field, its contribution to the redshift space
matter power spectrum
is the most complicated to model. We refer readers to Paper I for
details.

On the contrary, ${\bf v}_B$ has the smallest $O(1)$ Mpc correlation length, which is
shorter than scales of interest for RSD cosmology. This motivates us to treat it as an uncorrelated
field in our RSD modeling [Eq. (21), paper I]. Its impact on RSD is
completely captured by the damping function $D^{\rm FOG}_B$, which
will be quantified later in this paper.

Since a significant fraction of ${\bf v}_S$ comes from bulk motion,
${\bf v}_S$ has a  correlation length of $O(10)$ Mpc, larger than that of ${\bf v}_B$.
So we have to take into account its self-clustering [e.g., Eq. (25) of paper I].

\label{subsec:pdf}
\begin{figure*}[t]
\epsfxsize=18cm
\epsffile{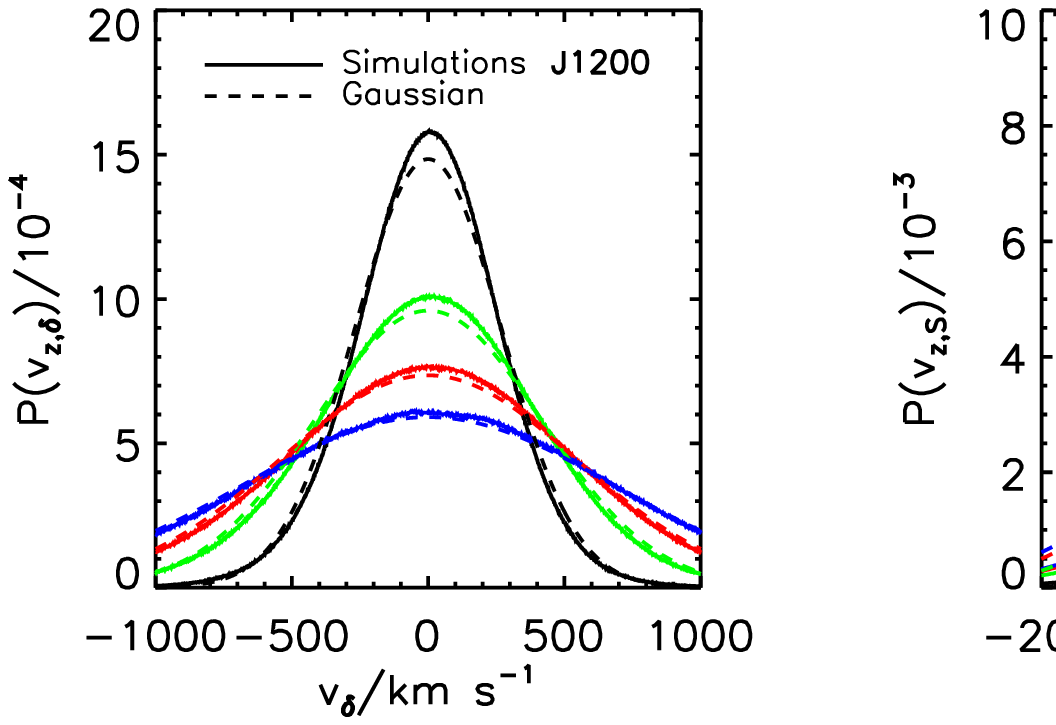}
\caption{The PDFs of ${\bf v}_{\delta,S,B}$
  along the $z$ axis are shown by solid lines.
  The dashed lines show Gaussian distributions with the same velocity mean and
  dispersion of corresponding velocity PDFs. Different line colors represent different
  redshifts. Apparently, ${\bf v}_\delta$ is the
  most Gaussian velocity component since it mainly correlates with linear matter
  density field and the window function $\tilde{W}$ suppresses
  non-Gaussianities from small scales. In contrast,  ${\bf v}_B$ is
  strongly non-Gaussian, consistent with the fact that
  most contribution comes from strongly nonlinear and non-Gaussian scales. }
\label{fig:pdf}
\end{figure*}
\begin{figure*}[t]
\epsfxsize=18cm
\epsffile{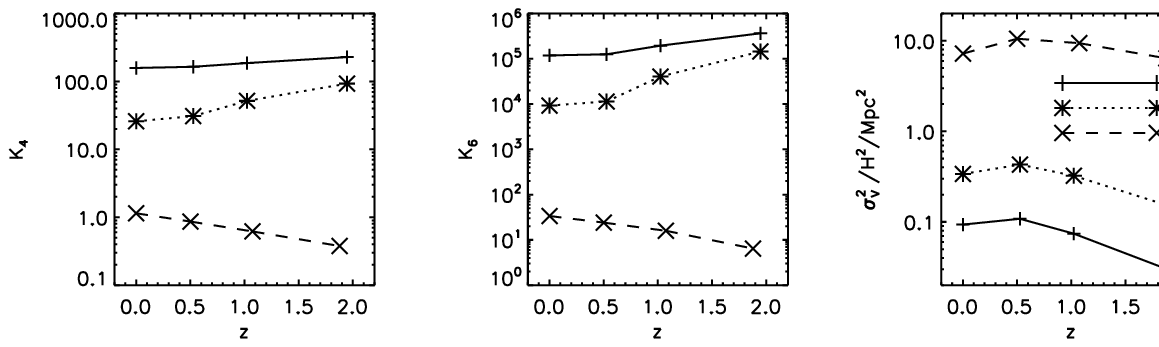}
\caption{
  \textit{Left and middle panels}: The fourth and sixth order of reduced cumulants of ${\bf
  v}_{\delta,S,B}$. They confirm the non-Gaussianity results shown in Fig. \ref{fig:pdf}.
  Towards lower redshift, non-Gaussianities of ${\bf v}_{S,B}$
  decrease, likely due to ongoing
  halo virialization and the associated velocity randomization, while non-Gaussianity
  of ${\bf v}_\delta$ increases due to nonlinear structure evolution.
  \textit{Right panel}: $\sigma_{v_\alpha}^2/H^2$ determines the
  leading-order  damping
  to redshift space clustering caused by the FOG effect
  [Eq. (\ref{eqn:DFOG}), Fig. \ref{fig:damping}].}
  \label{fig:cumul}
\end{figure*}

\subsection{The one-point velocity PDF and cumulants}
\label{subsec:PDF}
RSD modeling requires us to quantify the non-Gaussianity of the three
velocity components (Paper I). Fig. \ref{fig:pdf} shows $P(v_{z,\alpha})$
($\alpha=\delta,S,B$) at different redshifts. For
comparison, we also overplot the Gaussian distribution with the same
velocity mean (zero) and dispersion. Here,  $P(v_{z,\alpha})$ is the
corresponding PDF of the velocity component ${\bf v}_\alpha$ along the $z$-axis.

To better quantify the
non-Gaussianity, we calculate the reduced cumulants of the three velocity
components, $K_n$. The calculation is done against the real space velocity 
components on regular grid points. The real space velocity components are 
obtained by Inverse Fourier Transforming the Fourier space velocity components on 
regular gird points. Since the velocity field is symmetrical, $\left\langle
v^{2j+1}\right\rangle_c=0$. The nonvanishing cumulants are
\ba
\label{eqn:cumulants}
K_4&\equiv& \frac{\langle v^4\rangle}{\langle v^2\rangle^{2}}-3\no \ ,\\
K_6&\equiv& \frac{\langle v^6\rangle}{\langle
  v^2\rangle^{3}}-10\frac{\langle v^3\rangle^2}{\langle
  v^2\rangle^{3}}-15 \frac{\langle v^4\rangle}{\langle
  v^2\rangle^{2}}+30 \no\ ,\cdots
\ea
For Gaussian fields, $K_{n\geq3}=0$. But all three velocity components have
visible non-Gaussianity ( Fig. \ref{fig:cumul}).

\begin{figure*}[!htb]
\epsfxsize=15cm
\epsffile{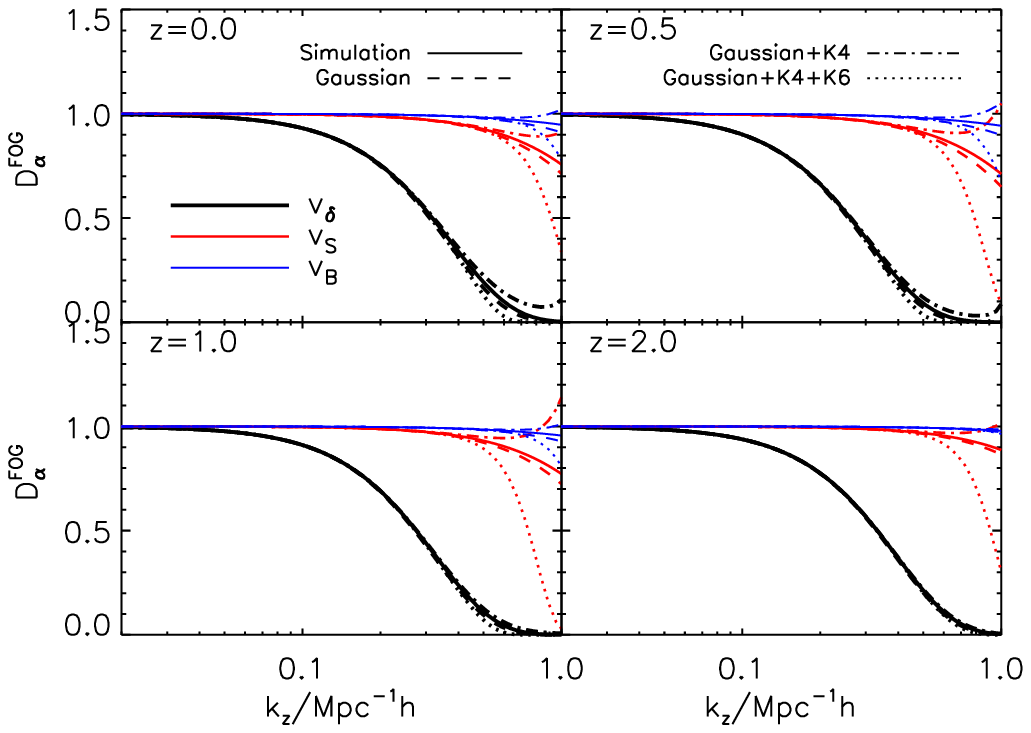}
\caption{
  The damping functions $D^{\rm FOG}_\alpha$.  They quantify the FOG
  effect of corresponding velocity components. The solid lines are
  calculated by Eq. (\ref{eqn:FOG}) through simulations. The other lines
  are approximations with/without $K_4$ and/or $K_6$
  terms in Eq. (\ref{eqn:cumulantexpansion}). The thick, intermediate thick, and thin lines
  with different colors represent results of $D^{\rm FOG}_\delta$,
  $D^{\rm FOG}_S$ and $D^{\rm FOG}_B$ respectively. Apparently ${\bf v}_\delta$ contributes most to
  the FOG effect since it has the largest velocity dispersion (right panel of Fig. \ref{fig:cumul}).
  The model tends to be more accurate towards higher redshift where nonlinearity/non-Gaussianity of
  matter/velocity field are smaller. The ratios between approximations and 
  simulation calculations are shown in Fig. \ref{fig:dampingratio}.}
\label{fig:damping}
\end{figure*}

\begin{figure*}[!htb]
\epsfxsize=15cm
\epsffile{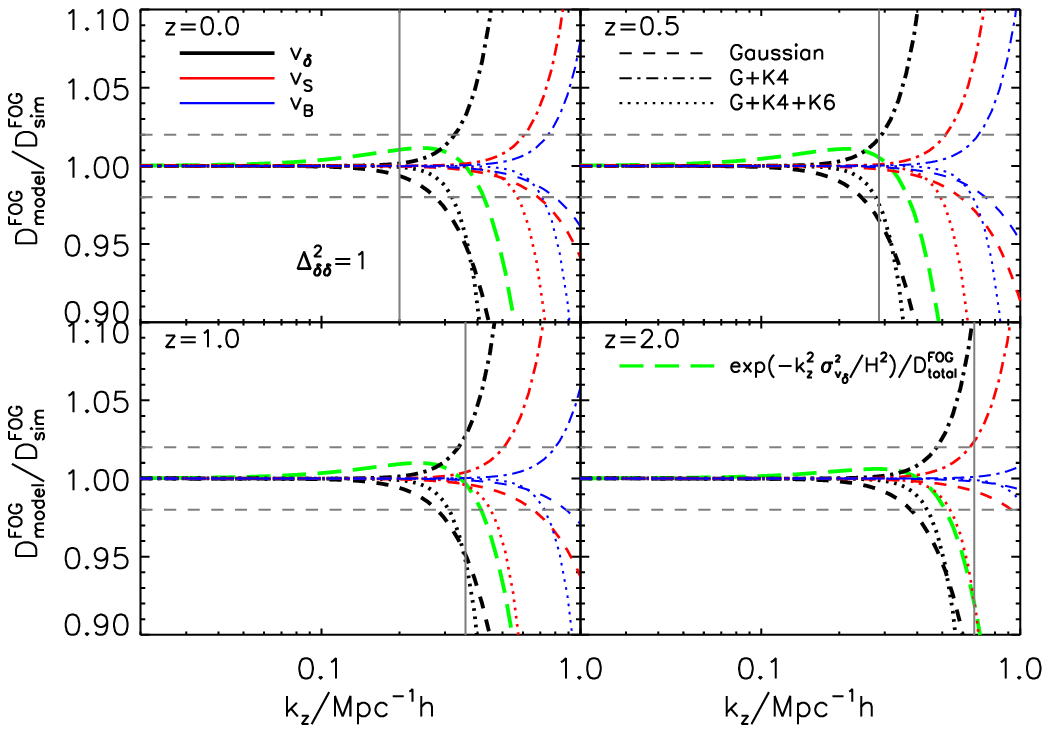}
\caption{
  The accuracy of various approximations of damping functions $D^{\rm FOG}_\alpha$.
  The dashed lines show that of the Gaussian approximation
  [Eq. (\ref{eqn:DFOG})]. The dot-dashed and 
  dotted lines show that of approximation with $K_4$ and $K_4+K_6$ terms respectively.
  The thick, intermediate thick, and thin lines with different colors
  correspond to $D^{\rm FOG}_\delta$,
  $D^{\rm FOG}_S$, and $D^{\rm FOG}_B$ respectively. The vertical gray
  lines denote the scale where $\Delta^2_{\delta\delta}(k,z)=1$. In
  particular, the thick green dashed lines verify that the approximation
  Eq. (\ref{eqn:deltaFOG}) is accurate  at $\sim 1\%$ level at
  $k<0.3h/$Mpc.} \label{fig:dampingratio} 
\end{figure*}

${\bf v}_\delta$ is nearly Gaussian, with $K_4\la 1$ at all
redshifts.  This is consistent with the fact that the dominant contribution comes from linear scales
(Fig. \ref{fig:velps}). However, it may appear to contradict the fact
that it is completely correlated with the density field, which is
highly nonlinear and non-Gaussian. The reason is that $\tilde{W}\ll 1$
at highly nonlinear scales, and hence it filters away most non-Gaussian
contribution.

In contrast,  ${\bf v}_B$ is strongly non-Gaussian (Fig. \ref{fig:cumul}),
consistent with the fact that most of its contribution comes from strongly
nonlinear and non-Gaussian scales. This result also supports our
conclusion that the aliasing effect is subdominant in ${\bf v}_B$
measured from the G100 simulation, otherwise we may expect a close-to-Gaussian ${\bf v}_B$.

The non-Gaussianity of ${\bf v}_S$
falls somewhere between, with visible departure from Gaussianity in
the PDF, $K_4>10$ and $K_6>10^4$.

An interesting behavior is that, $K_{4,6}$ of ${\bf v}_\delta$
increase towards lower redshift, while $K_{4,6}$ of ${\bf v}_S$ and
${\bf v}_B$ decrease (Fig. \ref{fig:cumul}). The increase of
non-Gaussianity in ${\bf v}_\delta$  indicates
that more and more ``Gaussian" scales have been converted to be
``non-Gaussian" due to the ongoing nonlinear evolution. On the other
hand, ongoing virialization in halos and the associated velocity
randomization may be responsible for decreasing non-Gaussianity in
${\bf v}_S$ and ${\bf v}_B$.

\subsection{The damping functions}
\label{subsec:damping}
Paper I proves that  all three velocity components contribute to the
FOG effect. Their contributions are described by the
corresponding damping function $D^{\rm FOG}_\alpha$ (Paper I),
\ba
\label{eqn:FOG}
\sqrt{D^{\rm FOG}_\alpha(k_z)}&\equiv &\left|\left\langle \exp\left(i\frac{k_zv_{z,\alpha}}{H}\right)
\right\rangle\right| \\
&=&\int_{-\infty}^{\infty} \exp\left(i\frac{k_zv_{z,\alpha}}{H}\right)
P(v_{z,\alpha})dv_{z,\alpha}\no \\
&=&\int _{-\infty}^{\infty}\cos\left(\frac{k_zv_{z,\alpha}}{H}\right)
P(v_{z,\alpha})dv_{z,\alpha} \ . \no
\ea

The cumulant expansion theorem allows us to express
$D^{FOG}_\alpha$ in terms of cumulants (Paper I),
\be
\label{eqn:cumulantexpansion}
\sqrt{D_\alpha^{\rm FOG}(k_z)}=\exp\left(-\frac{x}{2}\left[1-\frac{K_4}{12}x+\frac{K_6}{360}x^2+\cdots \right]\right)\ ,
\ee
where $x \equiv (k_z\sigma_{v_\alpha}/H)^2$. $\sigma_{v_\alpha}$ is the one-dimensional
velocity dispersion,
\ba
\label{eqn:sigmav}
\sigma^2_{v_\alpha}= \xi_{v_{z,\alpha}v_{z,\alpha}}(r=0)=\frac{1}{3}\int
\Delta^2_{v_\alpha v_\alpha}(k)\frac{dk}{k}\ .
\ea
To the first order, the damping functions take Gaussian form,
\be
\label{eqn:DFOG}
\sqrt{D_\alpha^{\rm
    FOG}(k_z)}=\exp\left(-\frac{x}{2}\right)=\exp\left(-\frac{(k_z\sigma_{v_\alpha})^2}{2H^2}\right)\ .
\ee
It is completely determined by $\sigma_{v_\alpha}/H$ (right panel, Fig. \ref{fig:cumul}).

We test the accuracy of Eq. (\ref{eqn:DFOG}) against the exact $D_\alpha^{\rm
  FOG}$, calculated from simulations using
Eq. (\ref{eqn:FOG}). Figures \ref{fig:damping} and \ref{fig:dampingratio}
show that Eq. (\ref{eqn:DFOG}) agrees well with the simulation for all three
velocity components, for a wide range of $k$. The accuracy is better
than $10\%$ where $\Delta^2_{\delta\delta}\la 1$.

Can adding higher-order terms such as $K_4$ and $K_6$ in
Eq. (\ref{eqn:cumulantexpansion}) improve the modeling accuracy of
$D_\alpha^{\rm FOG}$? This test is shown in Fig. \ref{fig:damping}
and Fig. \ref{fig:dampingratio}. Unfortunately, including
$K_4$ or $K_4+K_6$ does not necessarily improve the modeling accuracy of
$D^{\rm FOG}_\alpha$. Instead, including these terms often causes
unphysical behaviors such as $D_\alpha^{\rm FOG}>1$.

We thus conclude that the Gaussian approximation [Eq. (\ref{eqn:DFOG})] is in practice the optimal
approximation of $D_{\alpha}^{\rm FOG}(k_z)$, for all three velocity
components. Furthermore, since $\sigma_{v_\delta}^2\ga 10(
\sigma_{v_S}^2+\sigma_{v_B}^2)$, to high accuracy we can
approximate the overall damping function as
\ba
\label{eqn:deltaFOG}
D^{\rm FOG}(k_z)&\equiv &D_\delta^{\rm FOG}(k_z) D_S^{\rm FOG}(k_z) D_B^{\rm
  FOG}(k_z)\\
&\simeq&
\exp\left(-\frac{k^2_z(\sigma^2_{v_\delta}+\sigma^2_{v_S}+\sigma^2_{v_B})}{H^2}\right)\no
\\
&\simeq&  \exp\left(-\frac{k^2_z\sigma^2_{v_\delta}}{H^2}\right)\ . \no
\ea
Figure \ref{fig:dampingratio} verifies that the last approximation works
at $\sim 1\%$ accuracy at $k<0.3h/$Mpc and $z\in(0,2)$.

The excellent performance of the above Gaussian approximation is very
surprising, since much of the literature
prefers a Lorentz form $D^{\rm FOG}=1/(1+k_z^2\sigma_v^2/H^2)$
(e.g. \cite{Peacock94,Cole95,Diaferio96,Bromley97}) or more
complicated ones (e.g. \cite{kang05}). This seems to  contradict our findings. The point is that many of these studies infer the
damping function by fitting the form $P^s_{\delta\delta}(k,u)=P_{\delta\delta}(k)(1+fu^2)^2D^{\rm
  FOG}(ku)$ against simulations. However, there are ignored high-order corrections inside
of the parentheses [refer to Eq. (\ref{eqn:RD4}); For more details, refer
to Paper I]. 
Ignoring these corrections leads to misinterpretation of FOG. For
example, positive high-order corrections can be misinterpreted as a
damping function weaker than the Gaussian form. This issue will be
further clarified when we quantify the accuracy of the above RSD modeling
with simulations.

\section{Summaries and Discussions}
\label{sec:summary}
Numerical results presented in the current paper confirm many of our
qualitative arguments and speculations in Paper I. (1) They show that
${\bf v}_\delta$ dominates over ${\bf v}_S$ and ${\bf v}_B$ at $k<
0.5h/$Mpc. The ${\bf v}_\delta$ field is close to Gaussian, with a
correlation length of $O(100)$ Mpc. ${\bf v}_\delta$ has the largest
velocity one-dimension
dispersion, $\sigma_{v_\delta}\simeq270$ km/s at $z=0$.
It not only dominates the large-scale
enhancement of redshift space clustering, but also dominates the FOG
effect. We also measure a key function $\tilde{W}$ and confirm that it can indeed induce $O(10)\%$
underestimation in $f$.  We show by an example that $\tilde{W}$ has very limited
degrees of freedom and can be described by a simple fitting
formula accurately. (2) The ${\bf v}_B$ field is subdominant at $k<1h/$Mpc and is
negligible at $k<0.5h/$Mpc. It has the smallest velocity
dispersion. Due to observed numerical artifacts, we can only obtain
the upper limit $\sigma_{v_B}\leq 30$ km/s ($z=0$).
Furthermore, it has the shortest correlation length
and can be treated as a random field. Its impact on RSD is fully
captured by the damping function $D^{\rm FOG}_B(k_z)$. ${\bf v}_B$ is highly non-Gaussian. However,
due to its small amplitude, $D^{\rm FOG}_B(k_z)$ is well approximated by a Gaussian
form at scales of interest.
 (3) The ${\bf v}_S$ component is sub-dominant, but non-negligible, at
$k<0.5h/$Mpc. $\sigma_{v_S}\simeq60$ km/s at $z=0$. Its correlation length is of $O(10)$ Mpc, so we have to take its clustering into  RSD modeling.

There are still many open issues regarding the velocity statistics. Some of them will be addressed in
our future works. We just list a few of them here. One immediate question is the halo velocity
field. Once we understand it, we can understand the galaxy velocity
field with the help of halo model. This piece of information is
essential to model the galaxy RSD. Another issue is cosmological
dependences of these velocity statistics. An associated question is
the information budget in each velocity component, as a function of
redshift and scale.

\section{Acknowledgements}
We thank Jiawei Shao and Yu Yu for useful discussions.
This work was supported by the National Science Foundation of China
(Grants No. 11025316, No. 11121062, No. 10873035, No. 1133003, No. 10873027, No. 11121062, and No. 11233005), the National Basic
Research Program of China (973 Program) under Grant No. 2009CB24901 and  the
CAS/SAFEA
International Partnership Program for  Creative Research Teams (KJCX2-YW-T23).

\appendix
\section{Testing the NP method}
\label{sec:NPtest}


\bfi{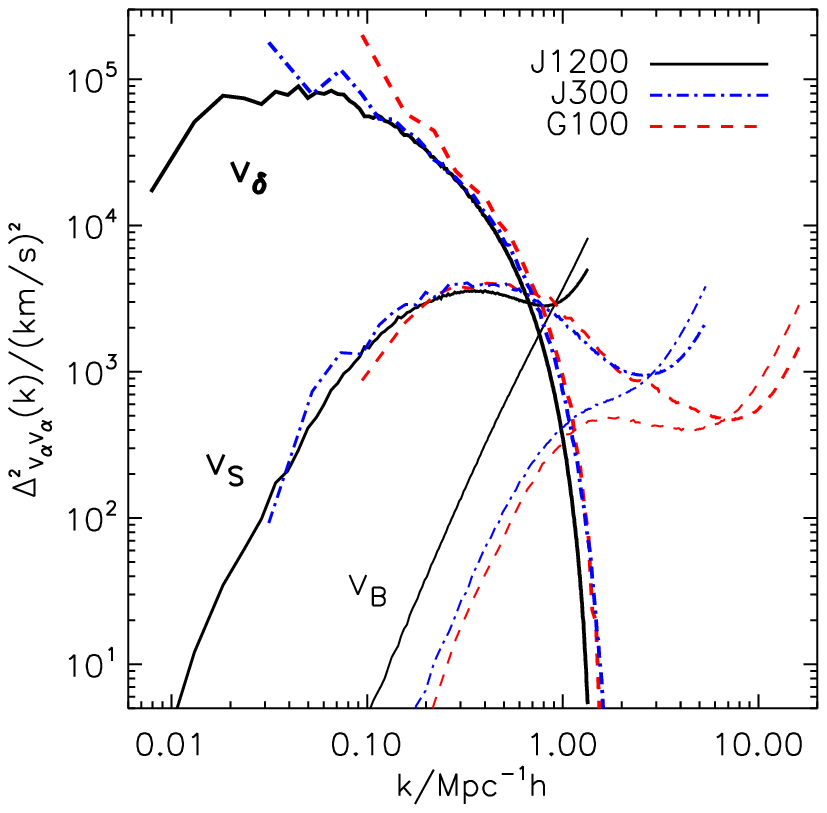}
\caption{Comparison between the three simulations at $z=0$. The velocity power
  spectra $\Delta^2_{v_\alpha v_\alpha}$  ($\alpha=\delta,S,B$) are calculated with
  $N_{\rm grid}=512^3$.  (1) ${\bf v}_\delta$ in J1200 loses
  power at $k\ga 0.3 h/$Mpc due to its low resolution, while ${\bf
    v}_\delta$ in G100 suffers from large cosmic variance at $k\la
  0.3h/$Mpc.  Combining the ${\bf v}_\delta$ measurement of J1200 at
  $k\la 0.3h/$Mpc and G100 at $k\ga 0.3h/$Mpc thus provides
  reliable determination of the ${\bf v}_\delta$ field.  Combining the
  J300 simulation at $k\sim 0.3h/$Mpc can be more reliable, although we
  have not done so in this paper. (2) The spurious increase of ${\bf
    v}_S$ at $k\sim 0.8h/$Mpc ($k\sim 2h/$Mpc)
  measured by J1200 (J300) is caused by numerical artifacts.   (3)
  Both J1200 and J300 fail to simulate ${\bf v}_B$ at any
  scales.
  }
\label{fig:velpsall}
\efi

The NP (Nearest-Particle) method we proposed is simple and straightforward to
implement. Despite its simplicity, we argue that it is robust in a
number of ways and is hence sufficiently accurate for the statistics
presented in this paper. We will run a number of convergence tests to
demonstrate its robustness. We will also quantify its accuracy for a
number of fiducial velocity fields, which resemble realistic velocity
fields.

\subsection{The convergence tests}
The convergence tests we consider are as follows: (1) Convergence
between the three simulations. J1200, J300
and G100 all have $1024^3$
particles, but have different box sizes and mass resolutions. By
comparing the three simulations, we  can estimate the reliable range
of the simulated velocity field.  (2) Convergence against the grid size.
We assign properties of simulation particles to regular grids in order to do
Ffast Fourier transformation (FFT). The price to pay is that information at
subgrid scales is smoothed out. Furthermore, as an approximated way of sampling
the velocity field, it can cause misidentification of different velocity components.
For example, finite grids are known to cause spurious ${\bf v}_B$ \cite{Pueblas09}.
By varying the grid number and checking for the convergence,
we can figure out suitable grid choices and a reliable range of the measured velocity statistics.
(3) {\it Sampling bias}. The issues addressed in (1) and (2) can be regarded as sampling biases.
But throughout the paper we refer to the sampling bias as that caused by the fact that we only
have velocity where there are particles. By sampling the velocity field using only a fraction
of particles, we amplify this sampling bias. Observing its dependence with respect to the fraction
of particles, we can estimate this sampling bias. In particular, if the measured velocity statistics
converge when the used fraction of particles is above a certain value, we will have reasonable
confidence on the velocity statistics using all particles.

We only show test results at $z=0$, where numerical artifacts are the
most severe. Furthermore, we mainly test the convergence in the power
spectrum. The convergence tests can be extended to other statistics.

\subsubsection{Comparison between the three simulations}
\label{subsec:massresolution}
Simulation box size determines the lower limit of reliable range of $k$, while
mass resolution determines the upper limit of reliable range of $k$.

Figure \ref{fig:velpsall} plots $\Delta^2_{v_\alpha v_\alpha}$ ($\alpha=\delta,S,B$) calculated
respectively from J1200, J300 and G100. It shows various spurious behaviors
and various discrepancies between these simulations. These are
manifests of  numerical artifacts. In particular, $\Delta^2_{v_Bv_B}$
decreases rapidly with increasing resolution from J1200 to
G100. This is consistent with the finding in \cite{Pueblas09}. Further
tests, especially that in Sec. \ref{subsec:rigoroustest}, will  identify
the origin of this numerical artifact.  Figure \ref{fig:velpsall} implies by naive
scaling that, simulations of Gpc box size and $10^{12}$ particles are
needed to fully control these numerical artifacts. Although a daunting
task,  such simulation is within the 
capability of state of art computation.

Nevertheless, J1200/J300/G100 do converge here and there, where we
expect the simulation results to be reliable.  Figure \ref{fig:velpsall} (and figures hereafter)
implies that, combining these simulations, it is feasible to reliably
measure the power spectrum of the three velocity fields in the range of interest ($k\in[0.01,1]h/$Mpc).
In this paper, we combine J1200 at $k<0.3h/$Mpc and G100 at $k>0.3
h/$Mpc to measure $\Delta^2_{v_\delta v_\delta}$ and  $\Delta^2_{v_S v_S}$. We use G100
to calculate $\Delta^2_{v_B v_B}$.

For other statistics such as the one-point PDFs, cumulants, and the
  damping functions, different scales mix and we have difficulty combining these simulations. To address this, we take a simplified approach. Since the
power of ${\bf v}_\delta$ peaks at $k\la 0.1h$/Mpc where J1200 is the
most reliable, we only use J1200 to measure statistics of ${\bf
  v}_\delta$.  Both the power of ${\bf v}_S$ and ${\bf v}_B$ peak at
$k\ga 0.3 h/$Mpc where G100 is most reliable, so we use G100 to
measure statistics of ${\bf v}_{S,B}$.

\subsubsection{Grid size}
\label{subsec:gridnum}

\begin{figure*}[!htb]
\epsfxsize=15cm
\epsffile{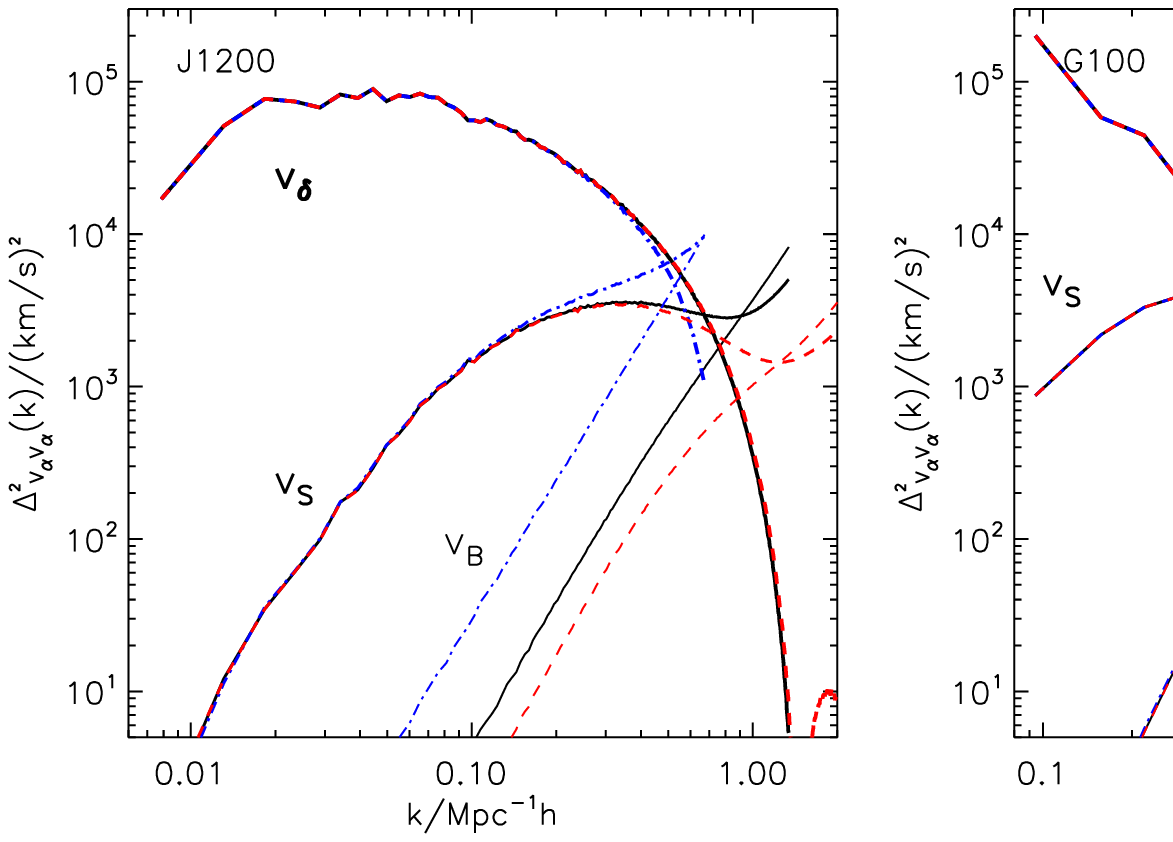}
\caption{Convergence tests on the grid size, for J1200 (left panel)
  and G100 (right panel) respectively. Fine grids are required to
  robustly sample the velocity and density field. For a given grid
  size, we should only trust the regimes where the result agrees with
  that of finer grids. Our tests show that $512^3$ grid of J1200 are
  needed to measure ${\bf v}_\delta$ and ${\bf v}_S$ at $k\la
  0.3h/$Mpc. Going to smaller scales should resort to the G100
  simulation. $512^3$ grids of G100 are needed to measure ${\bf
    v}_B$ at $k\la 1h/$Mpc.}
\label{fig:velpsgrid}
\end{figure*}

We calculate the power spectrum by assigning particle properties
(mass, velocity, etc.) onto regular gird points and performing
FFT. A natural convergence test is then against the grid
size $L_{\rm grid}$. Figure \ref{fig:velpsgrid} shows $\Delta^2_{v_{\alpha}v_\alpha}$
($\alpha=\delta,S,B$) calculated in the J1200 and G100 simulations
respectively, with $N_{\rm grid}=256^3,512^3,1024^3$. (1) Convergence tests
with J1200 show that grid-size-associated numerical artifacts in ${\bf
  v}_\delta$ are negligible at $k\la 0.3h/$Mpc, for grid size $L_{\rm grid}=4.7
h^{-1}$Mpc ($N_{\rm grid}=256^3$).  They are negligible at $k\la 1h/$Mpc,
for grid size $L_{\rm grid}=2.3 h^{-1}$Mpc ($N_{\rm grid}=512^3$).
Convergence tests with G100 show that numerical artifacts in ${\bf v}_\delta$
associated with grid size are negligible at $k\la 1h/$Mpc,
for grid size $L_{\rm grid}<0.4 h^{-1}$Mpc ($N_{\rm grid}=256^3$).
(2) J1200 and G100 show that grid-size-associated numerical artifacts
in ${\bf v}_S$ are
negligible at $k\times  {\rm Mpc/h}\la 0.1$ ($0.3$, $1$, $2$), for $L_{\rm
  grid}/h^{-1}{\rm Mpc}=4.7$ ($2.3$, $0.4$, $0.2$). (3) Convergence
tests with G100 show that, to reliably measure ${\bf v}_B$ to
$k=1h/$Mpc, grid size $L_{\rm grid}\leq 0.4h/$Mpc is required. To
reach $k= 2h/$Mpc, $L_{\rm grid}\leq 0.2h/$Mpc is required. We are
then able to conclude that, the measured power spectra shown in
Fig. \ref{fig:velps} are robust against numerical artifacts associated with
nonzero grid size (finite grid number).

Nonetheless, we do find significant disagreement between different
grid sizes over some $k$ ranges in Fig. \ref{fig:velpsgrid}. These are
clear manifestation of significant numerical artifacts associated with
grid size. The situation is the most severe for ${\bf v}_B$. No
convergence is found in J1200, meaning that even with $L_{\rm
  grid}=1.2h^{-1}$Mpc, it is not sufficient to accurately sample the ${\bf
  v}_B$ field. Tests with G100 show that $L_{\rm grid}\la 0.4h^{-1}$
Mpc is needed to accurately measure ${\bf v}_B$ at $k\la 1h/$Mpc.

There are at least three types of numerical artifacts associated with
grid size:  (1) One is the smoothing effect, suppression of small-scale power caused by the
assignment window function (e.g., \cite{Jing05} and references
therein). (2) Another is the alias effect caused by finite grid
number. It causes mixture of power among wave vectors
${\bf k}+2k_N{\bf n}$ \cite{Jing05}. Here $k_N\equiv \pi/L_{\rm grid}$
is the Nyquist wave number and $L_{\rm grid}$ is the grid
size. ${\bf n}$ is the three-dimensional integer vector. These two biases also exist
for the matter power spectrum measurement. (3) For the velocity
measurement, there is another type of alias effect.  It causes mixture
between different velocity components  \cite{Pueblas09}. This effect is especially
severe for ${\bf v}_B$ measurement.

All three types of bias depend
on the grid size (number).   The assignment window function for the NP
method is  $W_a({\bf r},{\bf r}_g)=\delta^D({\bf r}-{\bf r}_{NP}({\bf
  r}_g))$. Here ${\bf r}_g$ is the position of the given grid point
and ${\bf r}_{NP}({\bf r}_g)$ is the location of the corresponding
nearest particle. Notice that this window function is
inhomogeneous. Namely, it is not completely determined by ${\bf r}-{\bf
  r}_g$. So the formula of alias effect, derived with the condition $W_a({\bf r},{\bf r}_g)=W_a({\bf
  r}-{\bf r}_g)$ in \cite{Jing05}, does not apply here. It is beyond
the scope of this paper to derive a general expression of the alias
effect. Instead, we work on a limiting case of infinite particle number density
($N_P/L_{\rm box}^3\rightarrow \infty$). Under this limit,
${\bf r}_{NP}({\bf r}_g)\rightarrow {\bf r}_g$ and $W_a({\bf r},{\bf
  r}_g)\rightarrow \delta^D({\bf r}-{\bf
  r}_g)$. Following \cite{Jing05}, we obtain
\ba
\label{eqn:wa}
P^f_{\bf v\bf v}({\bf k})\rightarrow \sum_{\bf n}P_{\bf v\bf v}({\bf
  k}+2k_N{\bf n})\ .\label{eqn:NPpspecial}
\ea
Under this limit, the smoothing effect vanishes (the prefactor of the
power spectrum at ${\bf n}=(0,0,0)$ is unity). The alias effect, on one
hand, shows
as contaminations from ${\bf
  k}+2k_N{\bf n}$ modes with ${\bf n}\neq (0,0,0)$ to the measure
${\bf k}$ mode. On the other hand, it shows as leakages between
different velocity components (${\bf v}_\delta\rightarrow {\bf v}_S$,
${\bf v}_\delta\rightarrow {\bf v}_B$, ${\bf v}_S\leftrightarrow {\bf v}_B$).
 Sec. \ref{subsec:rigoroustest} will
quantify these leakages.  Both alias effects contribute to the decreasing of power at
small scales in both ${\bf v}_S$ and ${\bf v}_B$
(Fig. \ref{fig:velpsgrid}).

These convergence tests show that,  by combining J1200/J300/G100 we can safely neglect the
smoothing and aliasing caused by nonzero grid size at scale of
interest. So we do not attempt to correct for these numerical
artifacts. For methods of correcting them, refer to \cite{Jing05,Pueblas09,Koda13,Cui08}.

\subsubsection{Sampling bias}
\label{subsec:samplingbias}

\bfig{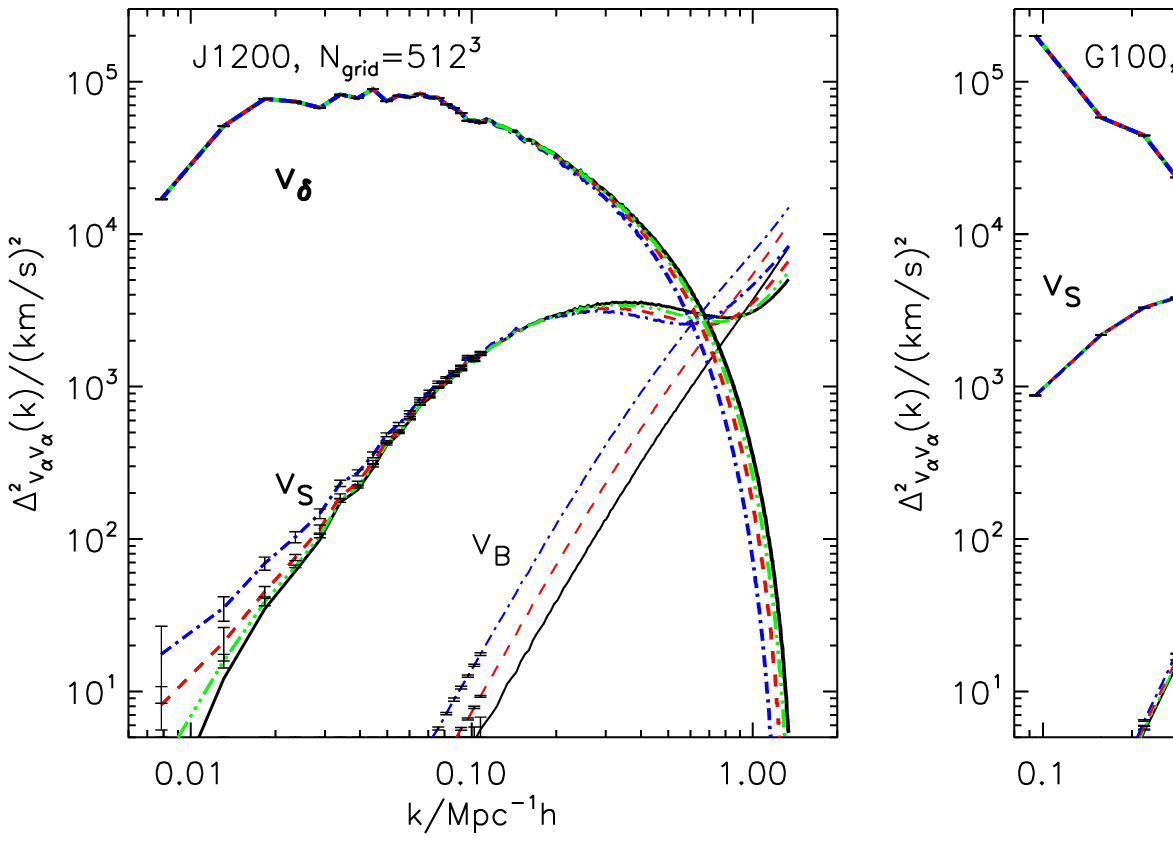}
\caption{$\Delta^2_{v_{\alpha}v_\alpha}$ ($\alpha=\delta,S,B$) of
  J1200 (left panel) and G100 (right panel)
  calculated using a fraction of randomly selected particles. This
  test quantifies the sampling bias.  
  The error bars are calculated by ten realizations of randomly
  selected $50\%$ and $25\%$ particles,
  and 20 realizations of $10\%$ particles. For clarity, not all error
  bars are shown. Test on J1200 finds visible sampling bias at
  $k\ga 0.3h/$Mpc in ${\bf v}_\delta$ and ${\bf v}_S$. It also tells
  us that ${\bf v}_B$ measured by J1200 is mainly noise. On the other
  hand, test on G100 shows much better convergence.}
\label{fig:testps}
\efig
\begin{figure*}[!htb]
\epsfxsize=18cm
\epsffile{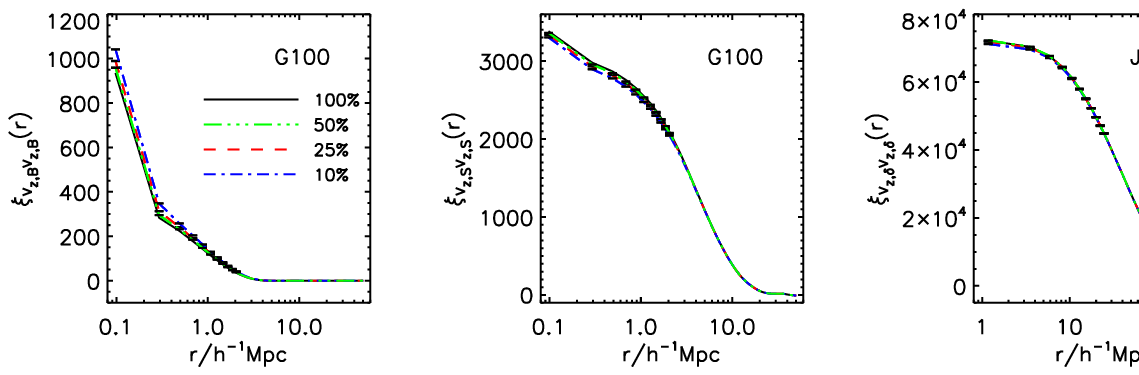}
\caption{The velocity correlation functions $\xi_{v_{z,\alpha}v_{z,\alpha}}$
  of three velocity components calculated
  using a fraction of randomly selected particles. The error
  bars are calculated by 10 realizations of randomly selected $50\%$ and $25\%$ particles,
  and 20 realizations of $10\%$ particles.
  The data point on the left end of each line is actually
  $\xi_{v_{z,\alpha}v_{z,\alpha}}(r=0)=\sigma^2_{v_\alpha}$.}
\label{fig:testcf}
\end{figure*}

\bfig{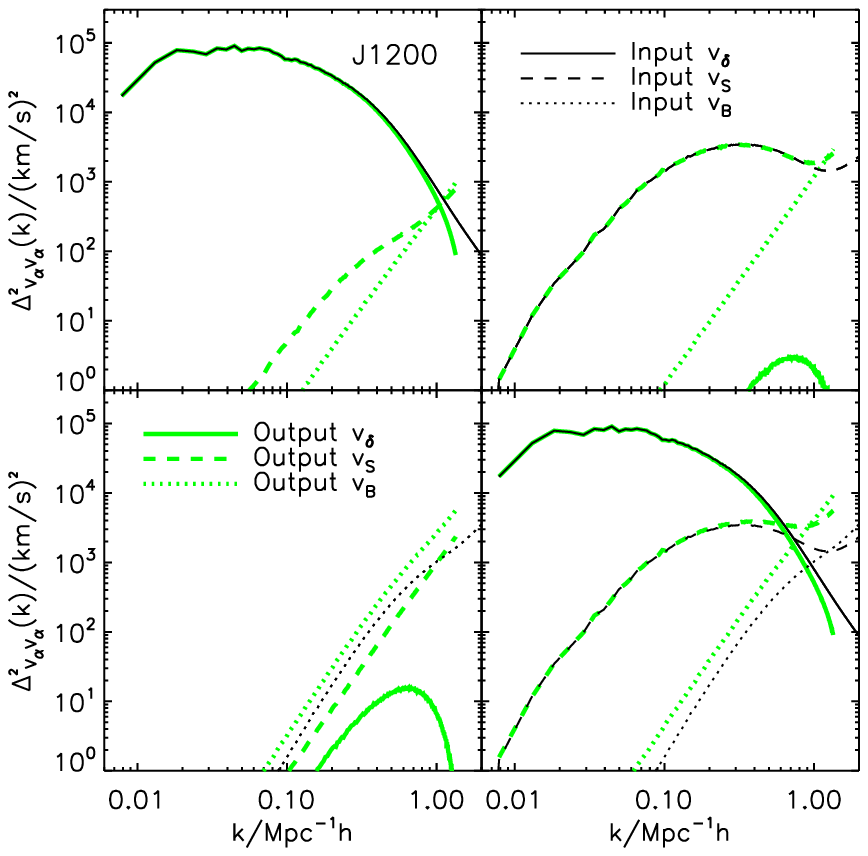}
\caption{Test against velocity fields of known statistics in J1200. The input velocity fields
  are constructed on $N_{\rm grid}=1024^3$.  The thin black solid, dashed and dotted lines represent input
  power spectra of ${\bf v}_\delta$, ${\bf v}_S$ and ${\bf v}_B$
  respectively.  We assign
  each simulation particle a velocity of its nearest grid point. We then apply
  the NP method against these ``simulations'' to measure the output
  power spectra of ${\bf v}_\delta$, ${\bf v}_S$ and ${\bf v}_B$,
  which are shown by thick green solid,
  dashed and dotted lines respectively. The measurement is performed on
  $N_{\rm grid}=512^3$ grid points.
  \textit{Top left panel} shows test case A:
  ${\bf v}_\delta\neq 0$ but ${\bf v}_S=0$ and ${\bf v}_B=0$.
  \textit{Top right panel} shows test case B:
  ${\bf v}_S\neq 0$ but ${\bf v}_\delta=0$ and ${\bf v}_B=0$.
  \textit{Bottom left panel} shows test case C:
  ${\bf v}_B\neq 0$ but ${\bf v}_\delta=0$ and ${\bf v}_S=0$.
  \textit{Bottom right panel} shows test case D:
  ${\bf v}_\delta\neq 0$, ${\bf v}_S\neq 0$ and ${\bf v}_B\neq
  0$. These tests clearly show leakages between any two velocity
  components and hence highlight and quantify these major numerical
  artifacts. They show that J1200 is robust to measure ${\bf
    v}_\delta$ and ${\bf v}_S$ at $k\la 0.3h/$Mpc, where leakages from
  other velocity components (${\bf v}_S\rightarrow {\bf v}_\delta$,
  ${\bf v}_B\rightarrow {\bf v}_\delta$, ${\bf v}_\delta\rightarrow
  {\bf v}_S$, ${\bf v}_B\rightarrow {\bf v}_S$) are insignificant. In contrast, leakages from ${\bf v}_\delta$ and ${\bf v}_S$ to ${\bf
  v}_B$ are too severe to measure ${\bf v}_B$ by J1200.  }
\label{fig:Rtest1J}
\efig


To highlight the sampling bias, we randomly select $50\%$, $25\%$, and $10\%$ of the
simulation particles and compare the measured velocity power spectra
with those of $100\%$ particles (Fig. \ref{fig:testps}). A complexity is that, the assignment window function of the NP method
$W_a({\bf r}, {\bf r}_g)$ is particle number density dependent. So
reducing the number of particles sampled also changes the alias
effect. Hence the numerical artifacts shown in Fig. \ref{fig:testps}
should be a mixture of sampling bias and alias effect. We do not
attempt to separate the two in this paper. 

For J1200, the power spectra of ${\bf
  v}_\delta$ and ${\bf v}_S$ converge at $k\la 0.2h/$Mpc. At
$k=0.3h/$Mpc, the results vary between the four cases ($10\%$, $25\%$, $50\%$,
and $100\%$). We then know that $1024^3\times 50\%$ particles
are not sufficient to simulate ${\bf v}_\delta$ and ${\bf v}_S$ at
$k\approx0.3h/$Mpc. However, we do not know if $1024^3$ particles are
sufficient. The situation for ${\bf v}_B$ is much worse. The power of
${\bf v}_B$ decreases with increasing number density at all
scales. This again confirms that the measured ${\bf v}_B$ in J1200 is
mainly numerical artifact, instead of real signal.

For G100, convergences are much better, since the sampling bias should decreases rapidly
towards higher particle number density. These convergences
tell us that sampling bias in a G100 like simulation with $1024^3$
particles and $100 h^{-1}$/Mpc box is negligible at $k\la 1h/$Mpc for
all the three velocity components.

Figure \ref{fig:testcf} shows correlation functions under the same
conditions.  It shows that the NP method gives considerably
accurate estimations for $\sigma^2_{v_\delta}$ and $\sigma^2_{v_S}$,
but overestimates $\sigma^2_{v_B}$. This is consistent with the large
discrepancies in $\Delta^2_{v_Bv_B}$ at large $k$ in
Fig. \ref{fig:testps}. It is hence challenging to accurately simulate
$\sigma_{v_B}$. A good thing is that, in our RSD modeling,
$\sigma_{v_B}$ can be chosen as  a free parameter to be fitted or even
set as zero. So the
inability of accurately simulating $\sigma_{v_B}$ is not a severe
problem.

\subsection{Testing against velocity fields of known statistics}
\label{subsec:rigoroustest}

\bfig{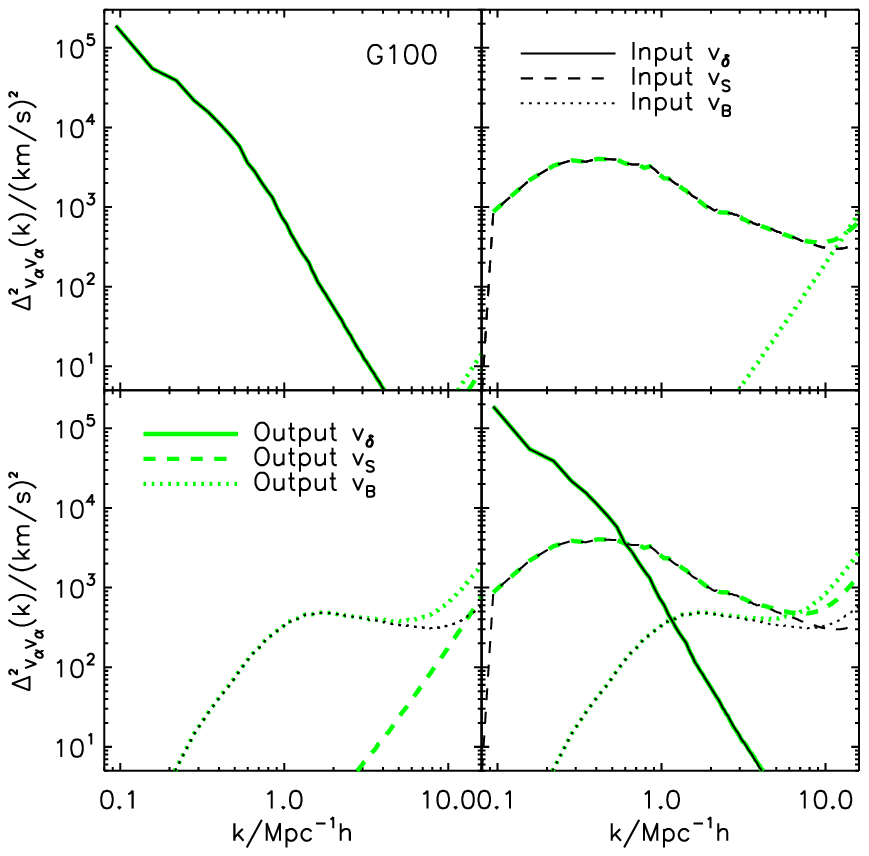}
\caption{Same as Fig. \ref{fig:Rtest1J} except here the test is done in G100.
  In the \textit{top left panel} the input and output
  $\Delta^2_{v_\delta v_\delta}$ almost overlap completely, meaning
  vanishing error in measuring ${\bf v}_\delta$ by G100. These tests
  also show that G100 is robust to measure ${\bf v}_S$ and ${\bf
    v}_B$ at $k\la 3h/$Mpc. Leakages between the three velocity
  components are all well controlled at $k\la 3h/$Mpc. }
\label{fig:Rtest1G}
\efig

The above convergence tests are good at diagnosing and setting up the
lower limit of some numerical artifacts. But they can be blind for
other artifacts and can have trouble disentangling various numerical
artifacts. A complementary test can be done with a fiducial velocity
field of known statistics (in our case the power spectrum). By
comparing the measured statistics with the input statistics, errors in
the measurement can be quantified straightforwardly. Obviously these
errors depend on the fiducial input. So the crucial step for this test
is the construction of the input velocity field as realistic as
possible. Our proposed velocity decomposition makes this step simpler
and more robust. This allows us to perform rigorous tests of the NP
method.

First we construct fiducial velocity fields of known statistics on
$N_{\rm grid}=1024^3$ regular grid
points. The fiducial ${\bf v}_\delta$ is constructed by a combination of measured density field and
the best fitted window function $W(k)$ in \S \ref{subsec:W}, $\theta_\delta({\bf k})=\delta(
{\bf k})W(k)$. The fiducial ${\bf v}_{S,B}$ can be generated using the
measured power spectra, if we assume Gaussianity. Alternatively, we
can directly use the measured ${\bf v}_{S,B}$ fields by the NP method
on regular $1024^3$ grid points from our simulations.  We will adopt
this second approach.

Secondly we assign each particle a velocity, which is the  velocity of
its nearest grid point. Along with the particle positions from
  the corresponding simulation (J1200/J300/G100), this is the new ``simulation'' we use to test
our NP method. We will assign these velocities to $N_{\rm grid}=512^3$
grid points to do the test. By setting some of the velocity components
to zero, we can generate up to seven velocity fields. We have tested the following
four cases of them. (A) ${\bf v}_\delta\neq 0$ but ${\bf v}_S=0$ and ${\bf v}_B=0$; (B)
${\bf v}_S\neq 0$ but  ${\bf v}_\delta=0$ and
${\bf v}_B=0$; (C) ${\bf v}_B\neq 0$, but ${\bf v}_\delta=0$ and ${\bf
  v}_S=0$; and (D) ${\bf v}_\delta\neq 0$, ${\bf v}_S\neq 0$ and ${\bf
  v}_B\neq 0$.

Tests on J1200 are shown in Fig. \ref{fig:Rtest1J}. (1) First we
  focus on the measurement accuracy in ${\bf v}_\delta$. Test on case
  A shows that the measured $\Delta^2_{v_{\delta}v_{\delta}}$
agrees well with the input one at $k\la 0.4h/$Mpc (top left panel,
Fig. \ref{fig:Rtest1J}). Tests on case B and C show that the leakages
from ${\bf v}_S$ and ${\bf v}_B$ to ${\bf v}_\delta$ are negligible.
So the measured ${\bf v}_\delta$ at $k\la 0.4 h/$Mpc is reliable. This
is consistent with the results of convergence tests.  (2) For ${\bf
  v}_S$, more careful interpretation should be given. Test on case B
shows that the measured $\Delta^2_{v_{S}v_{S}}$
agrees well with the input one at $k\la 0.8h/$Mpc (top right panel,
Fig. \ref{fig:Rtest1J}). However, tests on case A and case C show that the
leakage ${\bf v}_\delta\rightarrow {\bf v}_S$ and ${\bf
  v}_B\rightarrow {\bf v}_S$ are significant. Case D shows that we
should only trust the ${\bf v}_S$ measurement at $k\la 0.2
h/$Mpc. This is again consistent with findings through the convergence
test. (3) Test on case C shows that  the measured $\Delta^2_{v_B v_B}$
agrees with the input one, within a factor of 2 (bottom left panel,
Fig. \ref{fig:Rtest1J}). However, this result is very misleading. Tests
on case A and B shows that most of the output ${\bf v}_B$ is caused by the
leakage ${\bf v}_\delta\rightarrow {\bf v}_B$ and ${\bf
  v}_S\rightarrow {\bf v}_B$. In other words, ${\bf v}_B$ measured in
the J1200 simulation is mainly noise, instead of real signal. This is
consistent with our previous conclusion based on the convergence
tests.

Tests on G100 are shown in Fig. \ref{fig:Rtest1G}. By similar
argument, we can draw the conclusion that the NP method is accurate to
measure ${\bf v}_\delta$ and ${\bf v}_S$ at $k\sim 1h/$Mpc. Now the
measurement on ${\bf v}_B$ is significantly improved, since the leakages ${\bf v}_\delta\rightarrow {\bf v}_B$ and ${\bf
  v}_S\rightarrow {\bf v}_B$  are both subdominant at $k<1h/$Mpc.

These tests can be used to calibrate errors in the velocity
measurement. For example, by adding the power spectra of ${\bf v}_S$
and ${\bf v}_B$ in case A to that of ${\bf v}_\delta$ measured from
the real data, one can improve the measurement of the ${\bf v}_\delta$
power spectrum.

\bibliography{mybib}
\end{document}